\documentclass[11pt]{article}

\usepackage{graphicx,amssymb,amsbsy}
\usepackage{natbib}
\usepackage[colorlinks=true,linkcolor=blue,citecolor=blue]{hyperref}

\newcommand\Real{\mbox{Re}} % cf plain TeX's \Re and Reynolds number
\newsavebox{\astrutbox}
\sbox{\astrutbox}{\rule[-5pt]{0pt}{20pt}}

\def\pr{{\partial}}
\def\eps{{\epsilon}}

\def\nablap{\nabla_{\shortparallel}}

\begin{document}

\centerline{\Large{\bf Shallow-water models for a vibrating fluid}}

\vskip 5mm
\centerline{\bf Konstantin Ilin\footnote{Department of Mathematics, University of York,
Heslington, York YO10 5DD, UK. Email address for correspondence: konstantin.ilin@york.ac.uk}}

\begin{abstract}
We consider a layer of an inviscid fluid with free surface which is subject to vertical high-frequency vibrations. We derive
three asymptotic systems of equations that describe slowly evolving (in comparison with the vibration frequency) free-surface waves.
The first set of equations is obtained without assuming that the waves are long. These equations are as difficult to solve as the
exact equations for irrotational water waves in a non-vibrating fluid. The other two models describe long waves.
These models are obtained under two different assumptions about the amplitude of the vibration. Surprisingly, the governing equations
have exactly the same form in both cases (up to interpretation of some constants). These equations reduce to the standard dispersionless
shallow-water equations if the vibration is absent, and the vibration manifests itself via an additional term which makes the equations
dispersive and, for small-amplitude waves, is similar to the term that would appear if surface tension were taken into account.
We show that our dispersive shallow water equations have both solitary and periodic travelling waves solutions and discuss an analogy
between these solutions and travelling capillary-gravity waves in a non-vibrating fluid. 
\end{abstract}

%%%%%%%%%%%%%%%%%%%%%%%%%%%%%%%%%%%%%%%%%%%%%%%%%%%%%%%%%%%%%%%%%%%%%%%%%%%%%%%%%%%%%%%%%%%%%%%%%%%%%%%%%%%%%%%%%%%%%%%%%%%%%%%%%%%%%%%%%%%%%%%%%%%%%%

\setcounter{equation}{0}
\renewcommand{\theequation}{1.\arabic{equation}}

\section{Introduction}

It is well-known that high frequency vibrations of a tank containing a fluid with free surface or
two superimposed immiscible fluids can lead to
very interesting and non-trivial effects. For example, the Rayleigh-Taylor instability of two superimposed fluids
(with the heavier fluid on top of
the lighter one) can be
suppressed by vertical vibrations, and horizontal vibrations of the tank may lead to quasi-stationary finite-amplitude waves
on the interface \citep[see, e.g.,][]{Wolf1969, Wolf1970, Lyubimov2003}. Other examples of non-trivial effects of vibrations
include suppression of instability in liquid bridges \citep[]{Benilov2016},
parametric resonance (Faraday waves) \citep[e.g.][]{Miles, Vega2002}, steady streaming \citep[e.g.][]{Riley2001},
vibrational convection \citep[e.g.][]{Zen'kovskaya1966, Gershuni}, counterintuitive behaviour of solid particles
in a vibrating fluid \citep[e.g.][]{Sennitskii1985, Sennitskii1999, Sennitskii2007, Vladimirov2005} and
even a quantum-like behaviour of a droplet bouncing on the free surface
of a vibrating fluid \citep[see][]{Couder2005, Couder}.

In this paper, we consider an infinite horizontal fluid layer of finite depth which is subject to high-frequency vertical vibrations.
It is known that, under certain conditions, the dynamics of a periodically forced system can be described as a superposition of
a fast oscillatory motion and a slowly varying averaged motion. In this case, it is possible to obtain averaged equations describing
this slow evolution by employing a suitable averaging procedure \citep[see, e.g.,][]{Zen'kovskaya1966, Lyubimov2003, Yudovich2003}.
Here `slow' means that the characteristic time scale for these waves is much longer than the period of vibrations, i.e.
\begin{equation}
\omega \gg \sqrt{g/H} \label{0.1}
\end{equation}
where $\omega$ is the vibration frequency, $H$ in the mean fluid depth and $g$ is the gravitational acceleration.

For the flow regimes considered here to be observable, one needs to make sure that there is no parametric instability leading to
generation of Faraday waves \citep[see, e.g.,][]{Benjamin1954, Miles, Tuckerman1994, Vega2002}. The theory developed below
works in the limit of very high vibration frequency (much higher than the frequency range where the parametric instability usually occurs).
So, it will be assumed throughout the paper that either there is no parametric instability for some given values of the amplitude
and frequency of the vibration or the instability is suppressed by some other factor (e.g. by viscosity).

The aim of this paper is to derive and analyse nonlinear shallow water equations that describe slowly varying long waves on the surface
of a vertically vibrating layer of an inviscid fluid. Similar, but more general, equations without long wave approximation had been derived earlier by \citet[]{Lyubimov2003} and by \citet[]{Yudovich2003}. Somewhat similar averaged equations
had also been obtained for more complicated systems, which involve not only a free surface, but also some additional physical effects,
such as Marangoni effect \citep[see][and references therein]{Zen'kovskaya2007} or van der Waals forces between a rigid substrate and
a liquid film \citep[][]{Shklyaev2008, Shklyaev2009}. Here we focus on the pure effect of the vibration on free-surface flows. To make this effect as transparent as possible, we will consider the simplest problem and completely ignore compressibility, viscosity and surface tension. As far as we are aware, long-wave asymptotic behaviour of a vibrating fluid layer in this simple situation has not been considered before.

Let's briefly discuss  whether this simple problem can still be relevant for
real flows. The assumption that sound waves can be ignored means that
the typical hydrodynamic velocity is much
smaller than the speed of sound $c$, i.e.
\begin{equation}
H \omega \ll c .  \label{0.2}
\end{equation}
The viscosity can be dropped if the viscous time scale
is much greater than the typical period of the waves, i.e.
\begin{equation}
H^2/\nu \gg \sqrt{H/g} \label{0.3}
\end{equation}
where $\nu$ is the kinematic viscosity of the fluid.
Note that (\ref{0.1}) and (\ref{0.3}) imply that the thickness of viscous boundary layers is much less than the fluid depth:
$\sqrt{\nu/\omega} \ll H$. Combining (\ref{0.1}) and (\ref{0.2}), we obtain
\[
\sqrt{g/H} \ll \omega \ll c/H
\]
For water layer of depth $10$ cm, this is equivalent to
$9.9 \, \textrm{s}^{-1} \ll \omega \ll 1.48 \cdot 10^{4} \, \textrm{s}^{-1}$, which gives us quite a wide range of $\omega$ (say, from
$100 \, \textrm{s}^{-1}$ to $1000 \, \textrm{s}^{-1}$).
As was mentioned earlier, the effects of surface tension will not be considered for simplicity
(these effects can easily be taken into account later if necessary). This is a reasonable assumption
provided that the Bond number, defined as $Bo=\rho g L^2/\sigma$ (where $\rho$ is the fluid density, $L$ the typical wavelength and $\sigma$ the surface tension), is sufficiently large. For the water-air interface and the wavelength of $10$ cm, the Bond number is quite large ($Bo\approx 1.34\cdot 10^{3}$), so that the surface tension can be safely dropped.

The outline of the paper is as follows. Section 2 contains the formulation of the mathematical problem.
In section 3, we derive a general averaged model without the long-wave approximation (although
similar equations had been derived earlier by \citet[]{Lyubimov2003} and \citet[]{Yudovich2003}, we include this case for the sake of completeness and because our approach is different from that of the above papers). The asymptotic equations are Hamiltonian, and the dispersion
relation for small amplitude waves suggests that the effect of the vibration is similar to that of surface tension.
In section 4, we derive two long-wave asymptotic models: for the vibration amplitude much smaller than the fluid depth and for the vibration amplitude of the same order as the fluid depth. It turns out that these
two physically different situations lead to the same asymptotic equations. In section 5, we consider one-dimensional waves governed by the
equations derived in section 4. Here we show that the equations have travelling wave solutions in the form of both periodic and solitary waves. Finally,
section 6 contains a discussion of the results.

\begin{figure}
\begin{center}
\includegraphics*[height=7cm]{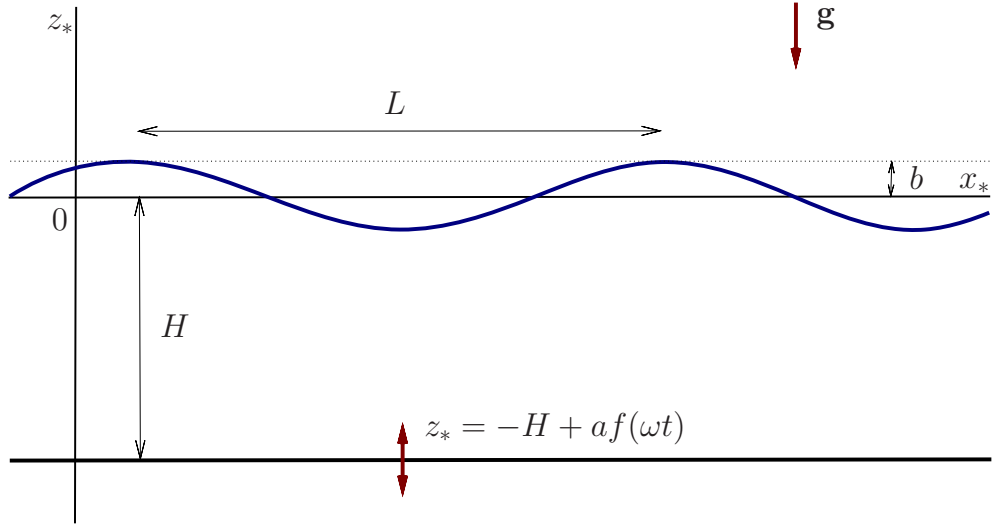}
\end{center}
\caption{Sketch of the flow.}
\label{wave_picture}
\end{figure}

%%%%%%%%%%%%%%%%%%%%%%%%%%%%%%%%%%%%%%%%%%%%%%%%%%%%%%%%%%%%%%%%%%%%%%%%%%%%%%%%%%%%%%%%%%%%%%%%%%%%%%%%%%%%%%%%%%%%%%%%%%%%%%%%%%%%%%%%%%%%%%%%%%%%%%

\setcounter{equation}{0}
\renewcommand{\theequation}{2.\arabic{equation}}

%%%%%%%%%%%%%%%%%%%%%%%%%%%%%%%%%%%%%%%%%%%%%%%%%%%%%%%%%%%%%%%%%%%%%%%%%%%%%%%%%%%%%%%%%%%%%%%%%%%%%%%%%%%%%%%%%%%%%%%%%%%%%%%%%%%%%%%%%%%%%%%%%%%%%%

\section{Basic equations}

\noindent
Consider an infinite layer of an inviscid flud over a flat rigid bottom which vibrates in vertical direction with amplitude $a$ and angular
frequency $\omega$ (see Fig. \ref{wave_picture}). Relative to the reference frame fixed in space, the equation of the bottom is $z_*= -H + a\, f(\omega t_*)$ where $z_*$ is
the vertical coordinate and $t_*$ is time.

\vskip 3mm
\noindent
In what follows we will work in the frame of reference vibrating with the bottom. Relative to it, the flow domain is
\[
D_*=\{(x_*,y_*,z_*)\in\mathbb{R}^3\vert -\infty < x_*,y_* < \infty, \ -H < z_* < \eta_*(x_*,y_*,t_*)\}
\]
where $x_*$, $y_*$ and $z_*$ are Cartesian coordinates; $z_*=\eta_*(x_*,y_*,t_*)$ is the equation of the free surface.
It is assumed that in the undisturbed state, $\eta_*(x_*,y_*,t_*)=0$.

\vskip 3mm
\noindent
Under the assumption that the flow is irrotational,
the equations of motion and boundary conditions can be written as
\begin{eqnarray}
&&\nabla_*^{2}\phi_*=0 \quad \hbox{in} \ \ D_*, \nonumber \\
&&\pr_{t_*}\phi_*+\frac{\vert\nabla_*\phi_*\vert^{2}}{2}+(g + a \ddot{f})\eta_*=0 \quad \hbox{at} \ \ z_*=\eta_*(x_*,y_*,t_*), \nonumber \\
&&\pr_{t_*}\eta_*+\pr_{x_*}\phi_{*}\pr_{x_*}\eta_*+\pr_{y_*}\phi_{*}\pr_{y_*}\eta_*
=\pr_{z_*}\phi_{*} \quad \hbox{at} \ \ z_*=\eta_*(x_*,y_*,t_*), \nonumber \\
&&\pr_{z_*}\phi_*=0 \quad \hbox{at} \ \ z_*=-H, \nonumber
\end{eqnarray}
where $\ddot{f}=d^2 f(\omega t_*)/dt_{*}^2$, $\nabla_*=(\pr_{x_*}, \pr_{y_*}, \pr_{z_*})$. We will assume that either $\phi_*$ and $\eta_*$ are periodic in $x_*$ and $y_*$ or some conditions at infinity are imposed (e.g., $\phi_*\to 0$ and $\eta_*\to 0$ as $\sqrt{x_{*}^2+y_{*}^2}\to \infty$).

\vskip 3mm
\noindent
Now we introduce the dimensionless variables $x, y, z, \tau, \phi, \eta$ defined as
\[
x_*=L x, \quad y_*=L y, \quad z_*=H z, \quad \tau=\omega t_*, \quad \phi_*=a \omega H \phi, \quad \eta_*=b \, \eta.
\]
Here $L$ is the characteristic length scale in horizontal direction, $H$ is the depth of the layer in the undisturbed state and $b$ is the characteristic scale for the displacement of the free surface from its undisturbed position.

\vskip 3mm
\noindent
In the dimensionless variables, the above equations take the form
\begin{eqnarray}
&&\phi_{zz} \, + \mu^2 \left(\phi_{xx}+\phi_{yy}\right) =0 \quad \hbox{in} \ \ D, \label{1} \\
&&\phi_{\tau}+\alpha \left( \mu^2 \, \frac{\phi_{x}^2+\phi_{y}^2}{2} + \frac{\phi_{z}^2}{2}\right)
+\beta \, [\gamma + f''(\tau)] \, \eta =0 \quad \hbox{at} \ \ z=\beta \, \eta(x,y,\tau). \label{2} \\
&&\eta_{\tau}+\alpha \mu^2\left(\phi_{x}\eta_{x}+\phi_{y}\eta_{y}\right) = \frac{\alpha}{\beta} \, \phi_{z} \quad \hbox{at} \ \ z=\beta \, \eta(x,y,\tau), \label{3} \\
&&\phi_{z} = 0 \quad \hbox{at} \ \ z=-1, \label{4}
\end{eqnarray}
where $D=\left\{(x,y,z)\in\mathbb{R}^3 \vert -\infty < x,y < \infty, \ -1 < z < \beta \, \eta(x,y,\tau)\right\}$; $\alpha$, $\beta$, $\gamma$ and
$\mu$  are dimensionless parameters defined as
\[
\alpha=\frac{a}{H}, \quad \beta=\frac{b}{H}, \quad \gamma=\frac{g}{a\omega^2}, \quad \mu=\frac{H}{L},
\]
so that $\alpha$ is the dimensionless amplitude of the vibrations, $\beta$ the dimensionless amplitude of the free surface waves,
$\gamma$ the ratio of the gravitational acceleration to the acceleration due to the vibrations and $\mu$ is the standard long-wave parameter
(the ratio of the fluid depth to the wavelength).

In what follows we deal with waves of finite amplitude corresponding to $\beta=1$. We will consider two cases: mean flows with $\mu=1$ and with $\mu\ll 1$ (long wave approximation).

%%%%%%%%%%%%%%%%%%%%%%%%%%%%%%%%%%%%%%%%%%%%%%%%%%%%%%%%%%%%%%%%%%%%%%%%%%%%%%%%%%%%%%%%%%%%%%%%%%%%%%%%%%%%%%%%%%%%%%%%%%%%%%%%%%%%%%%%%%%%%%%%%%%%%%

\setcounter{equation}{0}
\renewcommand{\theequation}{3.\arabic{equation}}

%%%%%%%%%%%%%%%%%%%%%%%%%%%%%%%%%%%%%%%%%%%%%%%%%%%%%%%%%%%%%%%%%%%%%%%%%%%%%%%%%%%%%%%%%%%%%%%%%%%%%%%%%%%%%%%%%%%%%%%%%%%%%%%%%%%%%%%%%%%%%%%%%%%%%%

%%%%%%%%%%%%%%%%%%%%%%%%%%%%%%%%%%%%%%%%%%%%%%%%%%%%%%%%%%%%%%%%%%%%%%%%%%%%%%%%%%%%%%%%%%%%%%%%%%%%%%%%%%%%%%%%%%%%%%%%%%%%%%%%%%%%

\section{Slow motions without long wave approximation}

%%%%%%%%%%%%%%%%%%%%%%%%%%%%%%%%%%%%%%%%%%%%%%%%%%%%%%%%%%%%%%%%%%%%%%%%%%%%%%%%%%%%%%%%%%%%%%%%%%%%%%%%%%%%%%%%%%%%%%%%%%%%%%%%%%%%

\noindent
Let $\mu=1$ and
\begin{equation}
\alpha=\eps, \quad \beta=1, \quad \gamma=\gamma_0 \eps \label{2.1}
\end{equation}
where $\gamma_0$ is a constant of order 1 (i.e. $\gamma_0=O(1)$ as $\eps\to 0$). These assumptions imply that (i) the amplitude of vibrations
of the bottom is small in comparison with the fluid depth, (ii) the amplitude of the waves may be of the same order of magnitude
as  the fluid depth, and (iii) the frequency of vibrations is sufficiently high, so that the acceleration due to vibrations is much greater
than the gravitational acceleration.

Equations (\ref{1})--(\ref{4}) become
\begin{eqnarray}
&&\nabla^2\phi=0 \quad \hbox{in} \ \ D, \label{2.2} \\
&&\phi_{\tau}+f''(\tau) \, \eta + \eps \left( \frac{\vert\nabla\phi\vert^{2}}{2}+ \gamma_0\eta \right)=0 \quad \hbox{at} \ \ z=\eta(x,y,\tau), \label{2.3} \\
&&\eta_{\tau}+\eps \, \nablap\phi\cdot\nablap\eta=\eps \, \phi_{z} \quad \hbox{at} \ \ z=\eta(x,y,\tau). \label{2.4} \\
&&\phi_{z}=0 \quad \hbox{at} \ \ z=-1. \label{2.5}
\end{eqnarray}
Here $\nabla=(\pr_x,\pr_y,\pr_z)$ is the gradient in three dimensions and $\nablap=(\pr_x,\pr_y)$ is the two-dimensional
gradient (parallel to the bottom).
As was mentioned above, these equations should be supplemented with an additional condition
which will be either a periodicity in variables $x$ and $y$
or a condition on the behaviour of the solution at infinity (as $\sqrt{x^2+y^2} \to \infty$).
This condition will be specified later (if it is needed).

\subsection{Derivation of the asymptotic equations}

\noindent
We are interested in the behaviour of solutions of Eqs. (\ref{2.2})--(\ref{2.5}) in the limit $\eps\to 0$. To construct an asymptotic expansion
of a solution, we employ the method of multiple scales \citep[e.g.][]{Nayfeh}
and assume that the expansion has a form
\begin{equation}
\phi=\phi_0(x,y,z,\tau,t)+ \eps \, \phi_1(x,y,z,\tau,t)+\dots, \quad
\eta=\eta_0(x,y,\tau,t)+ \eps \, \eta_1(x,y,\tau,t)+\dots \label{2.6}
\end{equation}
where
\[
t=\eps\tau
\]
is the slow time.
On substituting these in Eqs. (\ref{2.2})--(\ref{2.5}) and collecting terms of the same order in $\eps$, we obtain
\begin{eqnarray}
&&\nabla^2\phi_0=0 \quad \hbox{in} \ \ D_0, \label{2.7} \\
&&\phi_{0\tau}+  f''(\tau)\eta_0=0 \quad \hbox{at} \ \ z=\eta_0(x,y,\tau,t), \label{2.8} \\
&&\phi_{0z}=0 \quad \hbox{at} \ \ z=-1,  \label{2.9} \\
&&\eta_{0\tau}=0  \label{2.10}
\end{eqnarray}
at leading order and
\begin{eqnarray}
&&\nabla^2\phi_1=0 \quad \hbox{in} \ \ D_0, \label{2.11} \\
&&\phi_{1\tau}+  f''(\tau)\eta_1 +\phi_{0z\tau}\eta_1 + \phi_{0t}+\frac{\vert\nabla\phi_0\vert^2}{2}+
\gamma_0 \eta_0 =0 \quad \hbox{at} \ \ z=\eta_0(x,y,\tau,t), \label{2.12} \\
&&\phi_{1z}=0  \quad \hbox{at} \ \ z=-1. \label{2.13} \\
&&\eta_{1\tau}+\eta_{0 t}+\nablap\phi_{0}\cdot\nablap\eta_{0}=\phi_{0z}  \quad \hbox{at} \ \ z=\eta_0(x,y,\tau,t) \label{2.14}
\end{eqnarray}
at first order.
Here $D_0=\left\{(x,y,z)\in\mathbb{R}^3 \vert -\infty < x,y < \infty, \ -1 < z < {\eta}_{0}(x,y,\tau,t)\right\}$.

\vskip 3mm
\noindent
Throughout the paper, we will use the following fact: any bounded $2\pi$-periodic function $g(\tau)$
can be presented in the form
\[
g(\tau)=\overline{g}+\tilde{g}(\tau)
\]
where
\[
\overline{g}=\frac{1}{2\pi} \, \int\limits_{0}^{2\pi}g(\tau) \, d\tau
\]
is the averaged part of ${g}(\tau)$ and
$\tilde{g}(\tau)=g(\tau)-\overline{g}$ is its oscillatory part (having zero mean).

\vskip 3mm
\noindent
Consider now the leading order equations (\ref{2.7})--(\ref{2.10}). It follows from Eq. (\ref{2.10}) that
$\eta_{0}$ does not depend on the fast time, i.e.
\begin{equation}
\eta_0=\overline{\eta}_0(x,y,t). \label{2.15}
\end{equation}
Note that this equation implies that $D_0$ does not depend on the fast time $\tau$, i.e. $D_0$
is the domain corresponding to the averaged (over the period in $\tau$) position of the free surface.

\vskip 3mm
\noindent
Substituting (\ref{2.15}) into Eq. (\ref{2.8}) and separating the oscillatory part, we find that
\begin{equation}
\tilde{\phi}_{0}(x,y,z,t,\tau)= -   f'(\tau) \, \overline{\eta}_0(x,y,t) \quad \hbox{at} \ \ z=\overline{\eta}_0(x,y,t). \label{2.16}
\end{equation}
The oscillatory parts of (\ref{2.7}) and (\ref{2.9}) yield
\begin{equation}
\nabla^2\tilde{\phi}_0=0  \quad \hbox{in} \quad D_{0} \label{2.17}
\end{equation}
and
\begin{equation}
\tilde{\phi}_{0z}=0 \quad \hbox{at} \ \ z=-1. \label{2.18}
\end{equation}
If function $\overline{\eta}_0(x,y,t)$ were known, we would be able to find $\tilde{\phi}_0$ by solving the Laplace equation
(\ref{2.17}) subject to the boundary conditions (\ref{2.16}) and (\ref{2.18}) and the periodicity or decay condition in $x$ and $y$.

\vskip 3mm
\noindent
Consider now the first-order equations (\ref{2.11})--(\ref{2.14}). On averaging Eqs. (\ref{2.12}) and (\ref{2.14}), we obtain
\begin{eqnarray}
&&\pr_t\overline{\phi}_{0}+\overline{f''(\tau)\eta_1} +\overline{(\pr_{\tau}\pr_z\phi_{0})\eta_1} +
\frac{\overline{\vert\nabla\phi_0\vert^2}}{2}+ \gamma_0 \, \overline{\eta}_0 =0 \quad \hbox{at} \ \ z=\overline{\eta}_0(x,y,t), \label{2.19} \\
&&\pr_t\overline{\eta}_{0}+\nablap\overline{\phi}_0\cdot\nablap\overline{\eta}_{0}
=\pr_z\overline{\phi}_{0}  \quad \hbox{at} \ \ z=\overline{\eta}_0(x,y,t). \label{2.20}
\end{eqnarray}
Note that
\[
\overline{\vert\nabla\phi_0\vert^2}=\vert\nabla\overline{\phi}_0\vert^2 + \overline{\vert\nabla\tilde{\phi}_0\vert^2}, \quad
\overline{f''(\tau)\eta_1} = -\overline{f'(\tau)\tilde{\eta}_{1\tau}}
\]
and
\[
\overline{(\pr_{\tau}\pr_z\phi_{0})\eta_1}= - \overline{(\pr_z\tilde{\phi}_{0}) \pr_{\tau}\tilde{\eta}_{1}}.
\]
So, Eq. (\ref{2.19}) can be written as
\begin{equation}
\pr_t\overline{\phi}_{0}  +
\frac{\vert\nabla\overline{\phi}_0\vert^2}{2} + \gamma_0 \, \overline{\eta}_0 =G(x,y,t) \quad \hbox{at} \ \
z=\overline{\eta}_0(x,y,t), \label{2.21}
\end{equation}
where
\begin{equation}
G(x,y,t)= \overline{f'(\tau)\pr_{\tau}\tilde{\eta}_1} +
\overline{(\pr_z\tilde{\phi}_{0}) \pr_{\tau}\tilde{\eta}_1}-\frac{\overline{\vert\nabla\tilde{\phi}_0\vert^2}}{2}  \quad \hbox{at} \ \ z=\overline{\eta}_0(x,y,t). \label{2.22}
\end{equation}
Equation (\ref{2.21}) implicitly depends on both $\tilde{\phi}_0$ and $\tilde{\eta}_1$, which appear in Eq. (\ref{2.22}).
Function $\tilde{\eta}_1(x,y,\tau,t)$ can be found from the oscillatory part of Eq. (\ref{2.14}) that can be written as
\begin{equation}
\tilde{\eta}_{1\tau}+\nablap\tilde{\phi}_{0}\cdot\nablap\overline{\eta}_{0} =
\pr_z\tilde{\phi}_{0}  \quad \hbox{at} \ \ z=\overline{\eta}_0(x,y,t), \label{2.23}
\end{equation}
and this equation can be solved once $\tilde{\phi}_0$ is known.

It follows from Eqs. (\ref{2.16})--(\ref{2.18}) that $\tilde{\phi}_0(x,y,z,\tau,t)$ can be written in the form
\begin{equation}
\tilde{\phi}_0(x,y,z,\tau,t)= - \Phi(x,y,z,t)f'(\tau)  \label{2.24}
\end{equation}
where $\Phi$ is a solution of the following boundary-value problem:
\begin{eqnarray}
&&\nabla^2\Phi=0 \quad \hbox{for} \ \ -\infty < x,y < \infty, \ \ -1 < z < \overline{\eta}_0(x,y,t), \label{2.25} \\
&&\Phi_{z}=0   \quad \hbox{at} \ \ z=-1, \label{2.26} \\
&&\Phi = \overline{\eta}_0(x,y,t)  \quad \hbox{at} \ \ z=\overline{\eta}_0(x,y,t), \label{2.27}
\end{eqnarray}
subject to (yet unspecified) additional conditions in variables $x$ and $y$ (periodicity or decay at infinity). Suppose that we can solve problem
(\ref{2.25})--(\ref{2.27}) and let $\Phi(x,y,z,t)$ be its solution. Evidently, it will depend on $\overline{\eta}_0(x,y,t)$. In what follows, we
will use the following notation
\begin{equation}
\left.\Phi_z\right\vert_{z=\overline{\eta}_0} = \hat{Q}(x,y,t).  \label{2.28}
\end{equation}
It follows from  (\ref{2.23})--(\ref{2.28}) that
\begin{eqnarray}
&&\overline{f' \, \pr_{\tau}\tilde{\eta}_1}=-\overline{f'^2} \, \left(\hat{Q}-\left(1-\hat{Q}\right)
\left\vert\nablap\overline{\eta}_{0}\right\vert^2 \right) , \label{2.29} \\
&&\overline{(\pr_z\tilde{\phi}_{0}) \pr_{\tau}\tilde{\eta}_1} = \overline{f'^2} \, \left(\hat{Q}^2-
\hat{Q}\left(1-\hat{Q}\right) \left\vert\nablap\overline{\eta}_{0}\right\vert^2\right), \label{2.30} \\
&&\overline{\vert\nabla\tilde{\phi}_0\vert^2} = \overline{f'^2} \,
\left(\hat{Q}^2 + \left(1-\hat{Q}\right)^2 \left\vert\nablap\overline{\eta}_{0}\right\vert^2 \right). \label{2.31}
\end{eqnarray}
On substituting these into (\ref{2.22}), we obtain
\begin{equation}
G=\varkappa \left[\frac{\hat{Q}^2}{2} - \hat{Q} + \frac{1}{2}\,
\left(1-\hat{Q}\right)^2 \left\vert\nablap\overline{\eta}_{0}\right\vert^2 \right] \label{2.32}
\end{equation}
where
\begin{equation}
\varkappa = \overline{f'^2}. \label{2.33}
\end{equation}
Finally, averaging Eqs. (\ref{2.7}) and (\ref{2.9}), we get
\begin{eqnarray}
&&\nabla^2\overline{\phi}_0=0 \quad \hbox{in} \ \ D_0, \label{2.34} \\
&&\overline{\phi}_{0z}=0 \quad \hbox{at} \ \ z=-1.  \label{2.35}
\end{eqnarray}
Equations (\ref{2.34}), (\ref{2.35}), (\ref{2.20}) and (\ref{2.21}) with $G$ given by (\ref{2.32}) represent
a closed system of equations governing slow
evolution of waves on the free surface of a vibrating fluid.

\vskip 2mm
\noindent
\textbf{Remark 1.} If instead of $\gamma=\gamma_{0}\eps$, we consider
\[
\gamma=\Gamma \, \eps^2,
\]
where $\Gamma=O(1)$ as $\eps\to 0$, then the last term (containing $\gamma_0$) on the left side of Eq. (\ref{2.21}) will not be present
in the equation. This case corresponds to stronger vibrations when the gravitational acceleration is so small in comparison
with the vibrational acceleration that the gravity effect is negligible even for slow motions.

%%%%%%%%%%%%%%%%%%%%%%%%%%%%%%%%%%%%%%%%%%%%%%%%%%%%%%%%%%%%%%%%%%%%%%%%%%%%%%%%%%%%%%%%%%%%%%%%%%%%%%%%%%%%%%%%%%%%%%%%%%%%

\subsection{Some properties of the asymptotic equations}
Let
\[
\zeta = \overline{\eta}_{0}, \quad \psi = \overline{\phi}_{0}.
\]
Then the averaged equations derived above can be written as
\begin{eqnarray}
&&\psi_{t}  +
\frac{\vert\nabla{\psi}\vert^2}{2} + \gamma_0 \, {\zeta} =\varkappa \left[\frac{\hat{Q}^2}{2} - \hat{Q} + \frac{1}{2}\,
\left(1-\hat{Q}\right)^2 \left\vert\nablap{\zeta}\right\vert^2 \right]
 \quad \hbox{at} \ \
z={\zeta}(x,y,t), \label{2.36} \\
&&{\zeta}_{t}+\nablap\psi\cdot\nablap\zeta
={\psi}_z  \quad \hbox{at} \ \ z=\zeta(x,y,t), \label{2.37}
\end{eqnarray}
where $\psi$ satisfies
\begin{equation}
\nabla^2\psi=0 \quad \hbox{for} \ \ -1 < z < \zeta(x,y,t) \quad \hbox{and} \quad \psi_z\vert_{z=-1}=0 \label{2.38}
\end{equation}
and where
\begin{equation}
\hat{Q}=\left.\Phi_{z}\right\vert_{z=\zeta(x,y,t)} \label{2.39}
\end{equation}
and $\Phi(x,y,x,t)$ is the solution of the problem
\begin{eqnarray}
&&\nabla^2\Phi=0 \quad \hbox{for} \ \ -\infty < x,y < \infty, \ \ -1 < z < \zeta(x,y,t), \nonumber \\
&&\Phi_{z}\vert_{z=-1}=0,   \quad \Phi\vert_{z=\zeta(x,y,t)} = \zeta(x,y,t). \label{2.40}
\end{eqnarray}
If there is no vibration, Eqs. (\ref{2.36})--(\ref{2.38}) reduce to the standard system of equations for
irrotational water waves in a non-vibrating fluid. The vibration leads to the appearance of the extra term on
the right side of Eq. (\ref{2.36}).

Equations (\ref{2.36})--(\ref{2.40}) conserve the energy, given by
\begin{equation}
H=\int\limits_{\cal D} dx dy \int\limits_{-1}^{\zeta} dz \left[\frac{\vert\nabla\psi\vert^2}{2}+
\varkappa \, \frac{\vert\nabla\Phi\vert^2}{2}\right]
+\int\limits_{\cal D} dx dy \, \gamma_0 \, \frac{\zeta^2}{2}. \label{2.41}
\end{equation}
Here the domain of integration ${\cal D}$ is either the rectangle of periods, if periodic (in $x$ and $y$) solutions are considered, or the whole plane $\mathbb{R}^2$.
In the latter case, it is assumed that $\psi$, $\Phi$ and $\zeta$ decay at infinity so that the integrals in (\ref{2.41}) exist.

Equations (\ref{2.36}), (\ref{2.37}) can be written in Hamiltonian form. If, following
\citet{Zakharov1968} \citep[see also][]{Miles1981}, we introduce function
\begin{equation}
\chi(x,y,t) \equiv \psi(x,y,z,t)\vert_{z=\zeta(x,y,t)}, \label{2.42}
\end{equation}
then $\psi(x,y,z,t)$ is uniquely determined by $\chi(x,y,t)$ as a solution of the problem
\begin{eqnarray}
&&\nabla^2\psi=0 \quad \hbox{for} \ \ -\infty < x,y < \infty, \ \ -1 < z < \zeta(x,y,t), \nonumber \\
&&\psi_{z}\vert_{z=-1}=0,   \quad \psi\vert_{z=\zeta(x,y,t)} = \chi(x,y,t) \label{2.43}
\end{eqnarray}
supplemented with appropriate boundary conditions in variables $x$ and $y$.

It is also convenient to introduce the following notation
\[
\hat{N}(x,y,t)=\psi_z\vert_{z=\zeta(x,y,t)}
\]
where $\psi(x,y,z,t)$ is a solution of problem (\ref{2.43}). Note that $\hat{N}(x,y,t)$ is uniquely determined by
$\chi(x,y,t)$.

To rewrite Eqs. (\ref{2.36}), (\ref{2.37}) in terms of $\chi$ and $\zeta$, we first observe that
\begin{eqnarray}
&&\chi_{t}  = \psi_{t}\vert_{z=\zeta}+\hat{N}\zeta_t, \nonumber \\
&&\nablap\chi = \nablap\psi\vert_{z=\zeta} + \hat{N}\nablap\zeta. \nonumber
\end{eqnarray}
Then, we use these to eliminate $\psi_{t}\vert_{z=\zeta}$  and $\nablap\psi\vert_{z=\zeta}$ from Eqs. (\ref{2.36}), (\ref{2.37}). As a result, we obtain
\begin{eqnarray}
\chi_{t}  &=& \hat{N}\left[\hat{N} - \nablap \zeta\cdot \left(\nablap\chi - \hat{N}\nablap\zeta\right)\right]
-\frac{\hat{N}^2}{2} \nonumber \\
&&-\frac{\vert\nablap{\chi}-\hat{N}\nablap\zeta\vert^2}{2} - \gamma_0 \, {\zeta} + \varkappa \left[\frac{\hat{Q}^2}{2} - \hat{Q} + \frac{1}{2}\,
\left(1-\hat{Q}\right)^2 \left\vert\nablap{\zeta}\right\vert^2 \right], \label{2.44} \\
{\zeta}_{t} &=& \hat{N} - \nablap \zeta\cdot \left(\nablap\chi - \hat{N}\nablap\zeta\right). \label{2.45}
\end{eqnarray}
It can be verified by direct calculation that Eqs. (\ref{2.44})--(\ref{2.45}) are equivalent to the canonical Hamiltonian equations
\begin{equation}
\zeta_t = \frac{\delta H}{\delta \chi}, \quad \chi_t = -\frac{\delta H}{\delta \zeta}. \label{2.46}
\end{equation}

\vskip 2mm
\noindent
\textbf{Remark 2.} One can consider small amplitude waves and linearise Eqs. (\ref{2.36})--(\ref{2.40}). For waves in the form
\[
{\zeta}=\Real\left(\hat{\zeta}e^{i(\mathbf{k}\cdot\mathbf{x}- \omega t)}\right), \quad
\overline{\phi}_{0}=\Real\left(\hat{\psi}(z)e^{i(\mathbf{k}\cdot\mathbf{x} - \omega t)}\right),
\]
where $\mathbf{k}=(k_1,k_2)$, the dispersion relation has the form
\[
\omega^2(\mathbf{k}) = \vert\mathbf{k}\vert \, \tanh\left(\vert\mathbf{k}\vert\right)
\left[\gamma_0 + \varkappa \, \vert\mathbf{k}\vert \, \tanh\left(\vert\mathbf{k}\vert\right)\right].
\]
If the vibrational parameter $\varkappa$ is zero, this reduces to the standard dispersion relation for the
surface gravity waves on a fluid layer of finite depth. If function $\tanh\left(\vert\mathbf{k}\vert\right)$ that appears in
the square brackets were replaced by $\vert\mathbf{k}\vert$, the above dispersion relation would coincide with
the dispersion relation for gravity-capillary waves with $\varkappa$ playing the role of surface tension.
For long waves ($\vert\mathbf{k}\vert\ll 1$), the dispersion relation
reduces to $\omega^2(\mathbf{k}) \approx \vert\mathbf{k}\vert^2 \,
\left(\gamma_0 + \varkappa \, \vert\mathbf{k}\vert^2 \right)$,
which, again, is the same as the long-wave limit of the dispersion relation for gravity-capillary waves. Therefore, it is natural
to expect that
the effect of the vibration is similar to that of surface tension at least for sufficiently long waves.

%%%%%%%%%%%%%%%%%%%%%%%%%%%%%%%%%%%%%%%%%%%%%%%%%%%%%%%%%%%%%%%%%%%%%%%%%%%%%%%%%%%%%%%%%%%%%%%%%%%%%%%%%%%%%%%%%%%%%%%%%%%%%%%%%%%%%%%%%%%%%%%%%%%%%%

\setcounter{equation}{0}
\renewcommand{\theequation}{4.\arabic{equation}}

%%%%%%%%%%%%%%%%%%%%%%%%%%%%%%%%%%%%%%%%%%%%%%%%%%%%%%%%%%%%%%%%%%%%%%%%%%%%%%%%%%%%%%%%%%%%%%%%%%%%%%%%%%%%%%%%%%%%%%%%%%%%%%%%%%%%%%%%%%%%%%%%%%%%%%

%%%%%%%%%%%%%%%%%%%%%%%%%%%%%%%%%%%%%%%%%%%%%%%%%%%%%%%%%%%%%%%%%%%%%%%%%%%%%%%%%%%%%%%%%%%%%%%%%%%%%%%%%%%%%%%%%%%%%%%%%%%%%%%%%%%%

\section{Slow motions in the long wave approximation}

%%%%%%%%%%%%%%%%%%%%%%%%%%%%%%%%%%%%%%%%%%%%%%%%%%%%%%%%%%%%%%%%%%%%%%%%%%%%%%%%%%%%%%%%%%%%%%%%%%%%%%%%%%%%%%%%%%%%%%%%%%%%%%%%%%%%

\noindent
Consider now the situation when $\mu \ll 1$. We will derive two asymptotic models. In the first model, the amplitude of
the vibrations is assumed to be small compared with the fluid depth, while in the second, it is of the same order as the fluid depth.
We will see that, surprisingly, both assumptions lead to the same asymptotic equations.

\subsection{Small-amplitude vibrations}

We assume that (cf. Eq. (\ref{2.1}))
\begin{equation}
\mu=\eps^{1/2}, \quad \alpha=\eps, \quad \beta=1, \quad \gamma=\Gamma \, \eps^2 \label{4.1}
\end{equation}
where $\Gamma$ is a constant of order 1 (i.e. $\Gamma=O(1)$ as $\eps\to 0$). These assumptions mean that (i) the typical wavelength is
much larger than the fluid depth, (ii) the amplitude of vibrations
of the bottom is small compared with the fluid depth, (iii) the amplitude of the waves may be of the same order of magnitude
as  the fluid depth, and (iv) the frequency of vibrations is high enough for the acceleration due to vibrations to be much higher
than the gravitational acceleration. Substituting (\ref{4.1}) into Eqs. (\ref{1})--(\ref{4}), we obtain
\begin{eqnarray}
&&\phi_{zz} + \eps \, \nablap^2\phi =0 \quad \hbox{for} \ \ -1<z<\eta(x,y,\tau), \label{4.2} \\
&&\phi_{\tau}  + f''(\tau) \, \eta + \eps \, \frac{\phi_{z}^2}{2} +
\eps^2 \,  \left(\frac{\vert \nablap\phi\vert^2}{2} + \Gamma \, \eta\right)=0 \quad \hbox{at} \ \ z=\eta(x,y,\tau), \label{4.3} \\
&&\eta_{\tau}+\eps^2 \, \nablap\phi\cdot\nablap\eta = \eps \, \phi_{z} \quad \hbox{at} \ \ z=\eta(x,y,\tau), \label{4.4} \\
&&\phi_{z} = 0 \quad \hbox{at} \ \ z=-1, \label{4.5}
\end{eqnarray}
As before, we are interested in the asymptotic behaviour of solutions of Eqs. (\ref{4.2})--(\ref{4.5}) in the limit $\eps\to 0$. Again,
we assume that the asymptotic expansion
of a solution has a form
\[
\phi=\phi_0(x,y,z,\tau,t)+ \eps \, \phi_1(x,y,z,\tau,t)+\dots, \quad
\eta=\eta_0(x,y,\tau,t)+ \eps \, \eta_1(x,y,\tau,t)+\dots
\]
where
\[
t=\eps^2\tau
\]
is the slow time. Note that this slow time is different from that employed in the preceding section (it is much slower).
On substituting these in Eqs. (\ref{4.2})--(\ref{4.5}) and collecting terms of the same order in $\eps$, we obtain
at leading order,
\begin{eqnarray}
&&\phi_{0zz}=0 \quad \hbox{for} \ \ -1<z< \eta_0(x,y,\tau,t), \label{4.6} \\
&&\phi_{0\tau}+  f''(\tau)\eta_0=0 \quad \hbox{at} \ \ z=\eta_0(x,y,\tau,t), \label{4.7} \\
&&\phi_{0z}=0 \quad \hbox{at} \ \ z=-1,  \label{4.8} \\
&&\eta_{0\tau}=0,  \label{4.9}
\end{eqnarray}
at first order,
\begin{eqnarray}
&&\phi_{1zz} = - \nablap^2\phi_{0}  \quad \hbox{for} \ \ -1<z< \eta_0(x,y,\tau,t), \label{4.10} \\
&&\phi_{1\tau}+  f''(\tau)\eta_1 +\phi_{0z\tau}\eta_1 + \frac{\phi_{0z}^2}{2}=0 \quad \hbox{at} \ \ z=\eta_0(x,y,\tau,t), \label{4.11} \\
&&\phi_{1z}=0  \quad \hbox{at} \ \ z=-1, \label{4.12} \\
&&\eta_{1\tau}=\phi_{0z}  \quad \hbox{at} \ \ z=\eta_0(x,y,\tau,t), \label{4.13}
\end{eqnarray}
at second order,
\begin{eqnarray}
&&\phi_{2zz} = - \nablap^2\phi_{1}  \quad \hbox{for} \ \ -1<z< \eta_0(x,y,\tau,t), \label{4.14} \\
&&\phi_{2\tau}+  f''(\tau)\eta_2 +\phi_{1z\tau}\eta_1 + \phi_{0z\tau}\eta_2 + \phi_{0zz\tau} \, \frac{\eta_1^2}{2}
 \nonumber \\
&&\quad \ \ + \, \phi_{0z}\phi_{1z} + \phi_{0z}\phi_{0zz}\eta_1+\phi_{0t}+
\frac{\vert \nablap\phi_{0}\vert^2}{2} + \Gamma \, \eta_{0}=0 \quad \hbox{at} \ \ z=\eta_0(x,y,\tau,t), \quad \label{4.15} \\
&&\phi_{2z}=0  \quad \hbox{at} \ \ z=-1, \label{4.16} \\
&&\eta_{2\tau}+\eta_{0t}+\nablap\phi_{0}\cdot\nablap\eta_{0} =  \phi_{0zz}\eta_1 + \phi_{1z} \quad \hbox{at} \ \ z=\eta_0(x,y,\tau,t), \label{4.17}
\end{eqnarray}
etc.

\subsubsection{Leading order equations}

Consider the leading order equations (\ref{4.6})--(\ref{4.9}). It follows from (\ref{4.9})
that
\begin{equation}
\eta_0=\overline{\eta}_{0}(x,y,t). \label{4.18}
\end{equation}
This, in turn, implies that Eq. (\ref{4.7}) can be written as
\[
\tilde{\phi}_{0\tau}+  f''(\tau) \, \overline{\eta}_0=0 \quad \hbox{at} \ \ z=\overline{\eta}_0(x,y,t).
\]
Here we replaced ${\phi}_{0}$ with $\tilde{\phi}_{0}$ as the averaged part of ${\phi}_{0}$ produces zero contribution to
this equation. On integrating it in $\tau$, we find that
\begin{equation}
\tilde{\phi}_{0}= -f'(\tau) \, \overline{\eta}_0 \quad \hbox{at} \ \ z=\overline{\eta}_0(x,y,t). \label{4.19}
\end{equation}
Function $\tilde{\phi}_{0}(x,y,z,\tau,t)$ must also satisfy the equation
$\tilde{\phi}_{0zz}=0$ and
the boundary condition
$\tilde{\phi}_{0z}=0$ at $z=-1$ that
follow from Eqs. (\ref{4.6}) and (\ref{4.8}), respectively. These have a consequence that $\tilde{\phi}_{0}$ does not depend on $z$.
This fact and Eq. (\ref{4.19}) imply that
\begin{equation}
\tilde{\phi}_{0}= -f'(\tau) \, \overline{\eta}_0(x,y,t)  \label{4.20}
\end{equation}
for $-1< z < \overline{\eta}_0(x,y,t)$ and all $x$, $y$.

It also follows from Eqs. (\ref{4.6}) and (\ref{4.8}) (after averaging in $\tau$) that $\overline{\phi}_{0}(x,y,z,t)$ must satisfy
\[
\overline{\phi}_{0zz}=0 \quad \hbox{for} \ \ -1 <  z < \overline{\eta}_0(x,y,t)
\quad \hbox{and} \quad \left.\overline{\phi}_{0z}\right\vert_{z=-1}=0.
\]
These imply that $\overline{\phi}_{0}$ does not depend on $z$ as well, i.e.
\begin{equation}
\overline{\phi}_{0}=\overline{\phi}_{0}(x,y,t).  \label{4.21}
\end{equation}

\subsubsection{First order equations} From Eq. (\ref{4.13}) and the fact that ${\phi}_{0z}=0$ (which follows from (\ref{4.20}) and (\ref{4.21})),
we deduce that
$\eta_1$ does not depend on the fast time $\tau$, i.e.
\begin{equation}
\eta_1=\overline{\eta}_{1}(x,y,t).  \label{4.22}
\end{equation}
Taking into account Eq. (\ref{4.18}), we average Eqs. (\ref{4.10}) and (\ref{4.12}). This yields
\begin{eqnarray}
&&\overline{\phi}_{1zz} = - \nablap^2\overline{\phi}_{0}  \quad \hbox{for} \ \ -1<z< \overline{\eta}_0(x,y,t), \label{4.23} \\
&&\overline{\phi}_{1z}=0  \quad \hbox{at} \ \ z=-1, \label{4.24}
\end{eqnarray}
The most general solution of Eqs. (\ref{4.23}) and (\ref{4.24}) is given by
\begin{equation}
\overline{\phi}_{1}= -\left(\frac{z^2}{2}+z\right)\nablap^2\overline{\phi}_{0} +B(x,y,t)  \label{4.25}
\end{equation}
where $B(x,y,t)$ is an arbitrary function.

In view of Eqs. (\ref{4.18}), (\ref{4.20}), (\ref{4.21}) and (\ref{4.22}), Eq. (\ref{4.11}) reduces to
\[
\tilde{\phi}_{1\tau}+  f''(\tau) \, \overline{\eta}_1 =0 \quad \hbox{at} \ \ z=\overline{\eta}_{0}(x,y,t).
\]
or, equivalently,
\begin{equation}
\tilde{\phi}_{1} = -  f'(\tau) \, \overline{\eta}_1  \quad \hbox{at} \ \ z=\overline{\eta}_{0}(x,y,t).  \label{4.26}
\end{equation}
Separating the oscillatory part in Eq. (\ref{4.10}) and using (\ref{4.20}), we find that
\[
\tilde{\phi}_{1zz} =  f'(\tau) \, \nablap^2\overline{\eta}_{0}  \quad \hbox{for} \ \ 0 < z < \overline{\eta}_{0}(x,y,t).
\]
This should be solved subject to (\ref{4.26}) and the condition $\tilde{\phi}_{1z}\vert_{z=-1}=0$ (that follows from (\ref{4.12})). The solution is
given by
\begin{equation}
\tilde{\phi}_{1} = f'(\tau) \, \left(\frac{z^2-\overline{\eta}_{0}^2}{2} +z-\overline{\eta}_{0}\right)\,
\nablap^2\overline{\eta}_{0} -  f'(\tau) \, \overline{\eta}_1.  \label{4.27}
\end{equation}
Equations (\ref{4.22}), (\ref{4.25}) and (\ref{4.27}) are the only consequences of the first order equations (\ref{4.10})--(\ref{4.13}) that are needed in what follows.

\subsubsection{Second order equations}

Consider now the second order equations (\ref{4.14})--(\ref{4.17}). In view of (\ref{4.18}), (\ref{4.20}),
(\ref{4.21}) and (\ref{4.22}),
equations (\ref{4.15}) and (\ref{4.17}) simplify to
\begin{eqnarray}
&&\phi_{2\tau}+  f''(\tau)\eta_2 +\phi_{1z\tau}\overline{\eta}_1 + \phi_{0t}+
\frac{\vert\nablap\phi_{0}\vert^2}{2} + \Gamma \, \overline{\eta}_{0}=0 \quad \hbox{at} \ \ z=\overline{\eta}_0(x,y,t), \quad \label{4.28} \\
&&\eta_{2\tau}+\overline{\eta}_{0t}+\nablap\phi_{0}\cdot\nablap\overline{\eta}_{0} =  \phi_{1z} \quad \hbox{at} \ \ z=\overline{\eta}_0(x,y,t). \label{4.29}
\end{eqnarray}
Averaging these, we get
\begin{eqnarray}
&&\overline{\phi}_{0t}+
\frac{\overline{\vert\nablap\phi_{0}\vert^2}}{2} + \Gamma \, \overline{\eta}_{0} - \overline{f'(\tau)\eta_{2\tau}} =0 \quad
\hbox{at} \ \ z=\overline{\eta}_0(x,y,t), \quad \label{4.30} \\
&&\overline{\eta}_{0t}+\nablap\overline{\phi}_{0}\cdot\nablap\overline{\eta}_{0} =  \overline{\phi}_{1z} \quad \hbox{at} \ \ z=\overline{\eta}_0(x,y,t). \label{4.31}
\end{eqnarray}
The oscillatory part of Eq. (\ref{4.29}) simplifies to
\begin{equation}
\tilde{\eta}_{2\tau}+\nablap\tilde{\phi}_{0}\cdot\nablap\overline{\eta}_{0} =
\tilde{\phi}_{1z} \quad \hbox{at} \ \ z=\overline{\eta}_0(x,y,t).  \label{4.32}
\end{equation}
Further calculations yield
\begin{eqnarray}
\overline{\vert\nablap\phi_{0}\vert^2} &=& \vert\nablap\overline{\phi}_{0}\vert^2+\overline{\vert\nablap\tilde{\phi}_{0}\vert^2}=
\vert\nablap\overline{\phi}_{0}\vert^2 + \varkappa \, \vert\nablap\overline{\eta}_{0}\vert^2, \nonumber \\
\left.\overline{\phi}_{1z}\right\vert_{z=\overline{\eta}_{0}} &=&  -
\left(\overline{\eta}_{0}+1\right) \nablap^2\overline{\phi}_{0}, \nonumber \\
\overline{f'(\tau)\eta_{2\tau}}  &=&
\overline{f'(\tau)\left(\tilde{\phi}_{1z}-\nablap\tilde{\phi}_{0}\cdot\nablap\overline{\eta}_{0}\right)} \nonumber \\
 &=&  \varkappa \, \left[\left(1+\overline{\eta}_{0}\right) \nablap^2\overline{\eta}_{0}
 +\vert\nablap\overline{\eta}_{0}\vert^2\right], \nonumber
\end{eqnarray}
where $\varkappa$ is given by Eq. (\ref{2.33}) and where we have used (\ref{4.20}), (\ref{4.25}), (\ref{4.27}) and (\ref{4.32}).
Now we substitute these into
Eqs. (\ref{4.30}) and (\ref{4.31}). As a result, we get
\begin{eqnarray}
&&\overline{\phi}_{0t}+ \frac{\vert\nablap\overline{\phi}_{0}\vert^2}{2}
+\varkappa \, \frac{\vert\nablap\overline{\eta}_{0}\vert^2}{2}
+ \Gamma \, \overline{\eta}_{0} -  \varkappa \, \left[\left(1+\overline{\eta}_{0}\right) \nablap^2\overline{\eta}_{0}
 +\vert\nablap\overline{\eta}_{0}\vert^2\right] =0, \quad \label{4.33} \\
&&\overline{\eta}_{0t}+\nablap\overline{\phi}_{0}\cdot\nablap\overline{\eta}_{0}
+\left(1+\overline{\eta}_{0}\right)\nablap^2 \overline{\phi}_{0} = 0. \label{4.34}
\end{eqnarray}
Equations (\ref{4.33}) and (\ref{4.34}) represent the closed system of equations governing the slow evolution of the free surface.
To simplify the notation, let
\[
\psi=\overline{\phi}_{0}, \quad \zeta=\overline{\eta}_{0}.
\]
Then Eqs. (\ref{4.33}) and (\ref{4.34}) can be written as
\begin{eqnarray}
&&{\psi}_{t}+ \frac{\vert\nablap{\psi}\vert^2}{2}
+ \Gamma \, \zeta -  \varkappa \, \left[(1+\zeta) \nablap^2\zeta+\frac{\vert\nablap{\zeta}\vert^2}{2}\right] =0, \quad \label{4.35} \\
&&{\zeta}_{t}+\nablap\cdot\left[(1+\zeta)\nablap{\psi}\right]= 0. \label{4.36}
\end{eqnarray}
If $\varkappa=0$ (i.e. there is no vibration), then these equations reduce to the standard shallow water model.

\vskip 2mm
\noindent
\textbf{Remark 3.} For small-amplitude waves in the form
\[
{\zeta}=\Real\left(\hat{\zeta}e^{i(\mathbf{k}\cdot\mathbf{x}- \omega t)}\right), \quad
{\psi}=\Real\left(\hat{\psi}(z)e^{i(\mathbf{k}\cdot\mathbf{x} - \omega t)}\right),
\]
where $\mathbf{k}=(k_1,k_2)$, the dispersion relation has the form
\begin{equation}
\omega^2(\mathbf{k}) = \vert\mathbf{k}\vert^2 \,
\left(\Gamma + \varkappa \, \vert\mathbf{k}\vert^2 \right). \label{4.37}
\end{equation}
Note that (\ref{4.37}) coincides with the standard dispersion relation for gravity-capillary waves if $\varkappa$ is interpreted as surface tension (cf. Remark 1).

\vskip 2mm
\noindent
\textbf{Remark 4.} Equations (\ref{4.35}) and (\ref{4.36}) can also be obtained as a long wave approximation of
Eqs. (\ref{2.36}) and (\ref{2.37}) of Section 3. Indeed, it can be shown that if we re-scale variables in Eqs. (\ref{2.36}) and (\ref{2.37}) as
\[
x \rightarrow \delta^{1/2}x, \quad y \rightarrow \delta^{1/2}y, \quad \gamma_0 \rightarrow \Gamma \delta
\]
and construct a standard asymptotic expansion in the limit $\delta\to 0$, then, at leading order,
we will get Eqs. (\ref{4.35}) and (\ref{4.36}). So, Eqs. (\ref{4.35}) and (\ref{4.36}) can also be viewed as a result of two successive asymptotic approximations applied to the exact equations (\ref{1})--(\ref{4}): first, we let $\alpha\to 0$ and keep $\mu$ fixed and then we let
$\mu\to 0$.

%%%%%%%%%%%%%%%%%%%%%%%%%%%%%%%%%%%%%%%%%%%%%%%%%%%%%%%%%%%%%%%%%%%%%%%%%%%%%%%%%%%%%%%%%%%%%%%%%%%%%%%%%%%%%%%%%%%%%%%%%%%%%%%%%%%%%%%%%

\subsection{Vibrations of finite amplitude}

Now we will drop our earlier assumption that the amplitude of vibrations is small (compared with the fluid depth). Namely, we assume
that
(cf. Eq. (\ref{4.1}))
\begin{equation}
\mu=\eps^{1/2}, \quad \alpha=1, \quad \beta=1, \quad \gamma=\gamma_0 \, \eps \label{4.38}
\end{equation}
where $\gamma_0=O(1)$ as $\eps\to 0$. In other words, our assumptions are: (i) the typical wavelength is
much larger that the fluid depth, (ii) the amplitude of vibrations is of the same order of magnitude as
the fluid depth, (ii) the amplitude of the waves is of the same order of magnitude
as  the fluid depth, and (iii) the frequency of vibrations is high enough for the acceleration due to vibrations to be much higher
than the gravitational acceleration (but not as high as in Section 4.1, cf. Eq. (\ref{4.1})). Substituting (\ref{4.38}) into Eqs. (\ref{1})--(\ref{4}), we obtain
\begin{eqnarray}
&&\phi_{zz} + \eps \, \nablap^2\phi =0 \quad \hbox{for} \ \ -1<z<\eta(x,y,\tau), \label{4.39} \\
&&\phi_{\tau}  + f''(\tau) \, \eta + \frac{\phi_{z}^2}{2} +
\eps \,  \left(\frac{\vert \nablap\phi\vert^2}{2} + \gamma_0 \, \eta\right)=0 \quad \hbox{at} \ \ z=\eta(x,y,\tau), \label{4.40} \\
&&\phi_{z} = 0 \quad \hbox{at} \ \ z=-1, \label{4.41} \\
&&\eta_{\tau}+\eps \, \nablap\phi\cdot\nablap\eta =  \phi_{z} \quad \hbox{at} \ \ z=\eta(x,y,\tau), \label{4.42}
\end{eqnarray}
We seek an asymptotic expansion
of the solution in the form
\[
\phi=\phi_0(x,y,z,\tau,t)+ \eps \, \phi_1(x,y,z,\tau,t)+\dots, \quad
\eta=\eta_0(x,y,\tau,t)+ \eps \, \eta_1(x,y,\tau,t)+\dots
\]
where
\[
t=\eps\tau
\]
is the slow time. Note that this slow time is different from the slow time employed in Section 4.1, but the same as
the slow time of Section 3.
Substituting this expansion in Eqs. (\ref{4.39})--(\ref{4.42}) and collecting terms of the same order in $\eps$, we obtain
the following sequence of equations:

\noindent
at leading order,
\begin{eqnarray}
&&\phi_{0zz}=0 \quad \hbox{for} \ \ -1<z< \eta_0(x,y,\tau,t), \label{4.43} \\
&&\phi_{0\tau}+  f''(\tau)\eta_0+ \frac{\phi_{0z}^2}{2} =0 \quad \hbox{at} \ \ z=\eta_0(x,y,\tau,t), \label{4.44} \\
&&\phi_{0z}=0 \quad \hbox{at} \ \ z=-1,  \label{4.45} \\
&&\eta_{0\tau}=\phi_{0z} \quad \hbox{at} \ \ z=\eta_0(x,y,\tau,t);  \label{4.46}
\end{eqnarray}
at first order,
\begin{eqnarray}
&&\phi_{1zz} = - \nablap^2\phi_{0}  \quad \hbox{for} \ \ -1<z< \eta_0(x,y,\tau,t), \label{4.47} \\
&&\phi_{1\tau}+  f''(\tau)\eta_1 +\phi_{0z\tau}\eta_1 + \phi_{0z}\phi_{1z} \nonumber \\
&&\quad \ \ + \, \phi_{0z}\phi_{0zz}\eta_1 + \phi_{0t} +
\frac{\vert \nablap\phi_{0}\vert^2}{2} + \gamma_0 \, \eta_{0} = 0 \quad \hbox{at} \ \ z=\eta_0(x,y,\tau,t), \label{4.48} \\
&&\phi_{1z}=0  \quad \hbox{at} \ \ z=-1, \label{4.49} \\
&&\eta_{1\tau}+\eta_{0t}+ \nablap\phi_0\cdot\nablap\eta_0  =\phi_{1z}+ \phi_{0zz}\eta_1  \quad \hbox{at} \ \ z=\eta_0(x,y,\tau,t); \label{4.50}
\end{eqnarray}
etc.

\subsubsection{Leading order equations}

Equations (\ref{4.43}) and (\ref{4.45}) imply that $\phi_{0}$ does not depend on $z$, i.e.
\begin{equation}
\phi_{0}=\phi_{0}(x,y,\tau,t),   \label{4.51}
\end{equation}
which means that, at leading order, the vertical velocity is zero, and the horizontal velocity is homogeneous across the fluid layer. Now we deduce
from Eqs. (\ref{4.46}) and (\ref{4.51}) that $\eta$ does not depend on the fast time $\tau$:
\begin{equation}
\eta_{0}=\overline{\eta}_{0}(x,y,t),   \label{4.52}
\end{equation}
In view of Eqs. (\ref{4.51}) and (\ref{4.52}), Eq. (\ref{4.44}) reduces to
\[
\phi_{0\tau}+  f''(\tau)\overline{\eta}_0 =0 \quad \hbox{at} \ \ z=\overline{\eta}_0(x,y,t),
\]
from which we determine the oscillatory part of $\phi_{0}$ at $z=\overline{\eta}_0$:
\begin{equation}
\tilde{\phi}_{0} = - f'(\tau)\overline{\eta}_0 ,   \label{4.53}
\end{equation}

\subsubsection{First order equations}

We start with Eq. (\ref{4.47}). Its most general solution that satisfies the boundary condition (\ref{4.49}) is given by
\begin{equation}
\phi_{1} = - \left(\frac{z^2}{2}+z\right) \, \nablap^2 \phi_{0}(x,y,\tau,t) + B(x,y,\tau,t)    \label{4.54}
\end{equation}
for an arbitrary function $B$.

Consider now Eqs. (\ref{4.48}) and (\ref{4.50}). With the help of (\ref{4.51}) and (\ref{4.52}), these can be written as
\begin{eqnarray}
&&\phi_{1\tau}+  f''(\tau)\eta_1 + \phi_{0t} +
\frac{\vert \nablap\phi_{0}\vert^2}{2} + \gamma_0 \, \overline{\eta}_{0} = 0 \quad \hbox{at} \ \ z=\overline{\eta}_0(x,y,t), \label{4.55} \\
&&\eta_{1\tau}+\overline{\eta}_{0t}+ \nablap\phi_0\cdot\nablap\overline{\eta}_0  =\phi_{1z}  \quad \hbox{at} \ \ z=\overline{\eta}_0(x,y,t). \label{4.56}
\end{eqnarray}
Averaging yields
\begin{eqnarray}
&&\overline{\phi}_{0t} +\overline{f''(\tau)\eta_1} +
\frac{\overline{\vert \nablap\phi_{0}\vert^2}}{2} + \gamma_0 \, \overline{\eta}_{0} = 0
 \quad \hbox{at} \ \ z=\overline{\eta}_0(x,y,t), \label{4.57} \\
&&\overline{\eta}_{0t}+ \nablap\overline{\phi}_0\cdot\nablap\overline{\eta}_0
=\overline{\phi}_{1z}  \quad \hbox{at} \ \ z=\overline{\eta}_0(x,y,t) . \label{4.58}
\end{eqnarray}
The oscillatory part of (\ref{4.56}) gives us the equation
\[
\tilde{\eta}_{1\tau}+\nablap\tilde{\phi}_0\cdot\nablap\overline{\eta}_0  =\tilde{\phi}_{1z}  \quad \hbox{at} \ \ z=\overline{\eta}_0(x,y,t),
\]
which, with the help of (\ref{4.53}) and (\ref{4.54}), can be written as
\begin{equation}
\tilde{\eta}_{1\tau}= f'(\tau)\left[
\vert\nablap\overline{\eta}_0\vert^2 + (1+\overline{\eta}_0)\nablap^2\overline{\eta}_0\right]. \label{4.59}
\end{equation}
It follows from Eqs. (\ref{4.53}), (\ref{4.54}) and (\ref{4.59}) that
\begin{eqnarray}
\overline{\vert\nablap\phi_{0}\vert^2} &=& \vert\nablap\overline{\phi}_{0}\vert^2+\overline{\vert\nablap\tilde{\phi}_{0}\vert^2}=
\vert\nablap\overline{\phi}_{0}\vert^2 + \varkappa \, \vert\nablap\overline{\eta}_{0}\vert^2, \nonumber \\
\left.\overline{\phi}_{1z}\right\vert_{z=\overline{\eta}_{0}} &=&  -
\left(\overline{\eta}_{0}+1\right) \nablap^2\overline{\phi}_{0}, \nonumber \\
\overline{f''(\tau)\eta_{1}}  &=& -\overline{f'(\tau)\tilde{\eta}_{1\tau}}=
-\varkappa \, \left[\left(1+\overline{\eta}_{0}\right) \nablap^2\overline{\eta}_{0}
 +\vert\nablap\overline{\eta}_{0}\vert^2\right], \nonumber
\end{eqnarray}
Finally, substituting these into Eqs. (\ref{4.57}) and (\ref{4.58}), we obtain a closed system of averaged equations which can be written in the form
\begin{eqnarray}
&&{\psi}_{t}+ \frac{\vert\nablap{\psi}\vert^2}{2}
+ \gamma_0 \, \zeta -  \varkappa \, \left[(1+\zeta) \nablap^2\zeta+\frac{\vert\nablap{\zeta}\vert^2}{2}\right] =0, \quad \label{4.60} \\
&&{\zeta}_{t}+\nablap\cdot\left[(1+\zeta)\nablap{\psi}\right]= 0. \label{4.61}
\end{eqnarray}
where we have used the same notation as in Section 4.1: $\psi=\overline{\phi}_{0}$ and $\zeta=\overline{\eta}_{0}$.

Evidently, if we replace $\gamma_0$ in Eqs. (\ref{4.60}) by $\Gamma$, then Eqs. (\ref{4.60}) and (\ref{4.61}) become exactly
the same as Eqs. (\ref{4.35}) and (\ref{4.36}).

\vskip 2mm
\noindent
\textbf{Remark 5.} The two long-wave asymptotic models derived in this section correspond to physically different
situations, yet the asymptotic equations are the same. This fact looks surprising, but it is not coincidental. It turns out that
the same asymptotic equations arise
in many (physically) different situations. It can be shown using the asymptotic procedure of the present paper that the same leading-order equations
emerge from Eqs. (\ref{1})--(\ref{4}) under the following conditions
\[
\mu \ll 1, \quad \alpha \mu^2 \ll 1, \quad \gamma = \gamma_0 \alpha \mu^2, \quad \beta =O(1), \quad \gamma_0 =O(1).
\]
This implies that $\alpha$ can be large, provided that $\mu$ is sufficiently small. Physically this means that the amplitude of vibrations can be much greater than the fluid depth
provided that the waves considered are sufficiently long. For example, if we let $\alpha \mu^2 =\eps$ and $\mu=\eps^{m}$ for any natural number $m$
and $\eps\ll 1$, then all the above conditions hold for $\alpha=\eps^{1-2m} \gg 1$.

\vskip 2mm
\noindent
\textbf{Remark 6.} Equations (\ref{4.60}) and (\ref{4.61}) conserve the energy, given by
\begin{equation}
H=\int
(1+\zeta) \, \left(\frac{\vert \nablap\psi \vert^2}{2} + \gamma_0 \, \frac{\zeta^2}{2} + (1+\zeta) \,
\frac{\vert\nablap\psi\vert^2}{2}\right) \, dx \, dy. \label{4.62}
\end{equation}
It is easy to verify that Eqs. (\ref{4.60}) and (\ref{4.61}) are Hamiltonian:
\[
\zeta_t = \frac{\delta H}{\delta \psi}, \quad \psi_t = -\frac{\delta H}{\delta \zeta} .
\]
Note also that the above Hamiltonian, $H$, can be obtained from Eq. (\ref{2.41}) just by assuming that $\psi$ and $\Phi$ in (\ref{2.41}) do not depend
on the vertical coordinate $z$ (which corresponds to the long-wave approximation) and then integrating in $z$.

%%%%%%%%%%%%%%%%%%%%%%%%%%%%%%%%%%%%%%%%%%%%%%%%%%%%%%%%%%%%%%%%%%%%%%%%%%%%%%%%%%%%%%%%%%%%%%%%%%%%%%%%%%%%%%%%%%%%%%%%%%%%%%%%%%%%%%%%%%%%%%%%%%%%%%

\setcounter{equation}{0}
\renewcommand{\theequation}{5.\arabic{equation}}

%%%%%%%%%%%%%%%%%%%%%%%%%%%%%%%%%%%%%%%%%%%%%%%%%%%%%%%%%%%%%%%%%%%%%%%%%%%%%%%%%%%%%%%%%%%%%%%%%%%%%%%%%%%%%%%%%%%%%%%%%%%%%%%%%%%%%%%%%%%%%%%%%%%%%%

%%%%%%%%%%%%%%%%%%%%%%%%%%%%%%%%%%%%%%%%%%%%%%%%%%%%%%%%%%%%%%%%%%%%%%%%%%%%%%%%%%%%%%%%%%%%%%%%%%%%%%%%%%%%%%%%%%%%%%%%%%%%%%%%%%%%

\section{One-dimensional waves}

%%%%%%%%%%%%%%%%%%%%%%%%%%%%%%%%%%%%%%%%%%%%%%%%%%%%%%%%%%%%%%%%%%%%%%%%%%%%%%%%%%%%%%%%%%%%%%%%%%%%%%%%%%%%%%%%%%%%%%%%%%%%%%%%%%%%

\noindent
The aim of this section is to demonstrate that the averaged equations (\ref{4.60}) and (\ref{4.61}) have both solitary and periodic travelling
wave solutions. Here we consider one-dimensional waves and assume that $\psi$ and $\zeta$ do not depend on $y$. It is convenient to rewrite
the one-dimensional version of Eqs. (\ref{4.60}) and (\ref{4.61}) in terms of the velocity, $u$, and the total depth of the fluid, $h$,
defined by
\[
u(x,t)=\psi_{x}(x,t), \quad h(x,t)=1+\zeta(x,t).
\]
In terms of $u$ and $h$, we have
\begin{eqnarray}
&&u_{t}+ \left(\frac{u^2}{2}
+ \gamma_0 \, h -  \varkappa \, \left[h \, h_{xx} + \frac{h_x^2}{2}\right]\right)_x =0, \quad \label{5.1} \\
&&h_{t}+\left(h \, u\right)_x= 0. \label{5.2}
\end{eqnarray}
Equation (\ref{5.2}) represents conservation law of mass. Equation (\ref{5.1}) is associated with conservation of momentum: with
the help of (\ref{5.2}) it can be rewritten as
\begin{equation}
\left(h u\right)_{t}+ \left(h \, u^2 + \gamma_0 \, \frac{h^2}{2}
-  \varkappa \, h^2 \, h_{xx}\right)_x =0, \quad \label{5.3}
\end{equation}
which is precisely the momentum conservation law. Equations (\ref{5.1}) and (\ref{5.2}) also imply the conservation law of energy:
\begin{equation}
\mathcal{E}_{t}+ \mathcal{W}_x =0 \quad \label{5.4}
\end{equation}
where
\begin{eqnarray}
&&\mathcal{E} = h \, \frac{u^2}{2}
+ \gamma_0 \, \frac{h^2}{2} + \varkappa \, h \, \frac{h_{x}^2}{2},  \label{5.5} \\
&&\mathcal{W} = h u \, \frac{u^2}{2}
+ \gamma_0 \, u \, h^2 - \varkappa \, h u \left( h h_{xx} + \frac{h_{x}^2}{2}\right)
 - \varkappa \, h h_{x}h_{t} . \label{5.6}
\end{eqnarray}
Our conjecture is that there are no other conserved quantities, depending on $u$, $h$ and $h_x$ only,
but we did not attempt to prove this.

Now let's show that Eqs. (\ref{5.1}) and (\ref{5.2}) have travelling wave solutions.
We look for solutions of these equations in the form
\[
u(x,t)=U(s), \quad h(x,t)=H(s), \quad s\equiv x-ct
\]
where $c$ is a real parameter (its positive and negative values corresponds to waves travelling to the right and to the left, respectively).
Substituting these into Eqs. (\ref{5.1}) and (\ref{5.2}) and integrating once in $s$, we arrive at the following equations:
\begin{eqnarray}
&&-cU+ \frac{U^2}{2}
+ \gamma_0 \, H -  \varkappa \, \left[H \, H'' + \frac{H'^2}{2}\right] = C, \quad \label{5.7} \\
&&-cH + H \, U = B, \label{5.8}
\end{eqnarray}
where $C$ and $B$ are constants of integration. It follows from From Eq. (\ref{5.8}) that
\begin{equation}
U = \frac{B}{H}+c, \label{5.9}
\end{equation}
and we use this to eliminate $U$ from Eq. (\ref{5.7}). As a result, Eq. (\ref{5.7}) can be written as
\begin{equation}
m \left(H \, H''+\frac{H'^2}{2}\right)= \frac{\hat{B}^2}{2H^2}+H - \hat{C}  \label{5.10}
\end{equation}
where
\begin{equation}
m = \frac{\varkappa}{\gamma_0}, \quad \hat{B}=\frac{B}{\sqrt{\gamma_0}}, \quad \hat{C}=\frac{C}{\gamma_0}+\frac{\hat{c}^2}{2},
 \quad \hat{c}=\frac{c}{\sqrt{\gamma_0}}. \label{5.11}
\end{equation}
Equation (\ref{5.10}) is valid for any $\varkappa \geq 0$ and $\gamma_0 >0$ and  contains three free parameters ($m$, $\hat{B}$ and $\hat{C}$).
The case of $\gamma_0 =0$ (when the vibrations dominate over
the gravity in the averaged dynamics) will be treated separately. To reduce the number of parameters and transform
Eq. (\ref{5.10}) to a more convenient form, we introduce the following new dependent and independent variables:
\begin{equation}
H(s)=\frac{\hat{C}}{3} \, R^{2/3}(\sigma), \quad \sigma = m^{-1/2}\left(\frac{\hat{C}}{3}\right)^{-1/2} s.  \label{5.12}
\end{equation}
We are interested in solutions of (\ref{5.10}) that are positive and bounded ($0 < H(s) < \infty$). For such solutions, $R(\sigma)$ is
well-defined, positive and bounded.
Equation (\ref{5.10}) can be rewritten in term of $R$ and $\sigma$ as
\begin{equation}
R''(\sigma) = - \hat{V}'(R)  \label{5.13}
\end{equation}
where
\begin{equation}
\hat{V}(R) = \frac{9}{8} \, \left( 4 q \, R^{-2/3} - R^{4/3} + 6 \, R^{2/3}\right)  \label{5.14}
\end{equation}
and
\begin{equation}
q = \frac{\hat{B}^2}{4} \, \left(\frac{\hat{C}}{3}\right)^{-3}. \label{5.15}
\end{equation}
Evidently, Eq. (\ref{5.13}) coincides with the equation of motion of a particle of unit mass moving in the potential $\hat{V}(R)$, with $\sigma$ playing
the role of time. Now it is easy to show the existence of travelling periodic and solitary wave solutions of Eqs. (\ref{5.1}) and (\ref{5.2}):
all we need to do is to find bounded periodic and non-periodic solutions of (\ref{5.13}). Note that, since only
the square of $\hat{c}$ appears in Eq. (\ref{5.10}), there are waves of the same shape travelling to the left and to the right,
and it is sufficient to consider only positive $\hat{c}$.

\begin{figure}
\begin{center}
\includegraphics*[height=10cm]{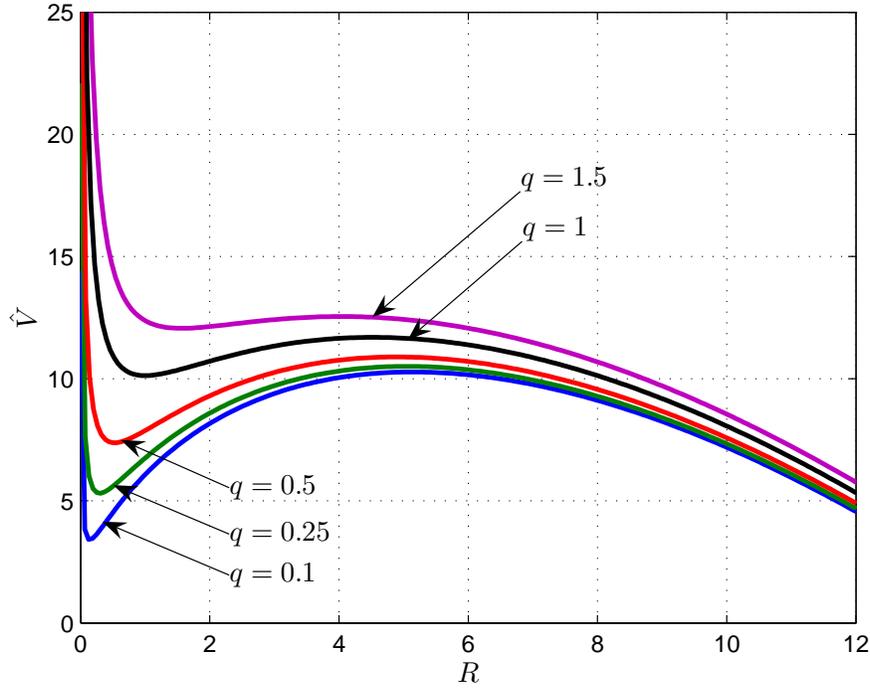}
\end{center}
\caption{Plots of $\hat{V}(R)$ for several values of parameter $q$ ($0<q<2$).}
\label{potential}
\end{figure}

The potential $\hat{V}(R)$ contains only one free parameter, $q$.
Its graphs for several values of $q$ are shown
in Fig. \ref{potential}.  For $0< q < 2$, function $\hat{V}(R)$ has a local minimum
at $R=R_1$ and a local maximum at $R=R_2$ where
\begin{equation}
R_1=\left[1-2\cos\left(\alpha/3 + \pi/3\right)\right]^{3/2}, \qquad
R_2=\left[1+2\cos\left(\alpha/3\right)\right]^{3/2}. \qquad
\cos\alpha=1-q.  \label{5.16}
\end{equation}
For $q<0$ or $q>2$, there are no local maximum and minimum, and, therefore, no solutions satisfying $0 < R(\sigma) < \infty$ for all $\sigma\in\mathbb{R}$ exist.

Below we separately consider solitary waves, periodic waves and the case of $\gamma_0=0$.

\subsection{Solitary waves}

Solitary travelling waves correspond to finite, but non-periodic motions of the particle.
In view of Fig. \ref{potential}, these requirements are satisfied if
the particle's energy, $\hat{E} = R'^{2}/2 + \hat{V}(R)$, is equal to the value of $\hat{V}$ at the local maximum, i.e.
$\hat{E} = \hat{V}(R_2)$. Thus, for each value of parameter $q\in(0,2)$, we obtain a solitary wave solution of
Eqs. (\ref{5.1}) and (\ref{5.2}). In what follows, we restrict our attention to solitary waves satisfying the conditions
\[
u(x,t) \to 0, \quad h(x,t)\to 1 \quad \hbox{as} \ \ \vert x\vert \to \infty,
\]
which, in terms of $H(s)$ and $U(s)$, have the form
\begin{equation}
U(s) \to 0, \quad H(s)\to 1 \quad \hbox{as} \ \ \vert s\vert \to \infty. \label{5.17}
\end{equation}
When $\sigma\to \pm\infty$, the particle approaches the point $R=R_2$. This fact and Eq. (\ref{5.17})
imply that
\begin{equation}
\frac{\hat{C}}{3}\left(1+2\cos\frac{\alpha}{3}\right)=1. \label{5.18}
\end{equation}
It also follows from conditions (\ref{5.17}) and Eqs. (\ref{5.7}) and (\ref{5.8}) that
\begin{equation}
B=-c, \qquad C=\gamma_0, \qquad \hat{B}= -\hat{c}, \qquad
\hat{C}= 1 + \frac{\hat{c}^2}{2}. \label{5.19}
\end{equation}
These and Eq. (\ref{5.15}) give us the following relationship between parameter $q$ and the wave speed $\hat{c}$:
\begin{equation}
q= \frac{27}{4} \, \hat{c}^2 \, \left( 1 + \frac{\hat{c}^2}{2}\right)^{-3}  . \label{5.20}
\end{equation}
Note that Eqs. (\ref{5.18}) and (\ref{5.20}) are not independent: if Eqs. (\ref{5.20}) holds, so does Eq. (\ref{5.18}), or \emph{vice versa}.
The requirement that $0<q<2$ and Eq. (\ref{5.20}) imply that $0 < \hat{c} < 1$, which means solitary wave can propagate only with speed less than
the phase speed of small-amplitude gravity waves, $\sqrt{\gamma_0}$.
\begin{figure}
\begin{center}
\includegraphics*[height=10cm]{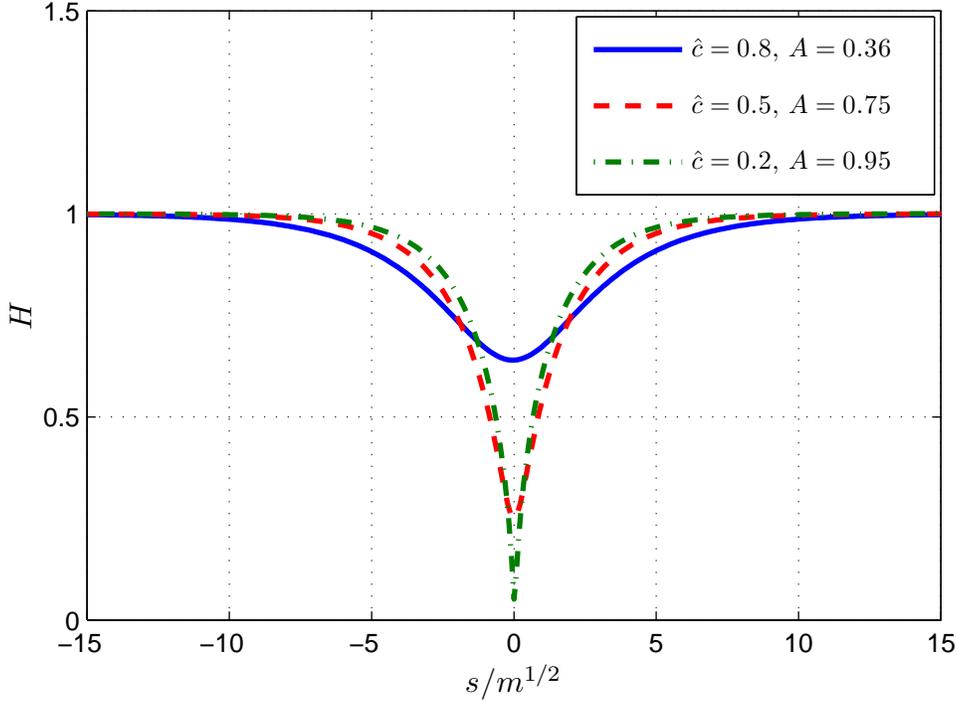}
\end{center}
\caption{Typical profiles of solitary waves.}
\label{sol_wave_fig}
\end{figure}

Equation (\ref{5.13}) was solved numerically for various values of $q$ using MATLAB built-in ODE solvers. Then, Eqn. (\ref{5.12}) were employed to compute $H(s)$ and the wave amplitude, $A$, defined as
\begin{equation}
A= \max_{s\in\mathbb{R}} \left\vert H(s)-1\right\vert. \label{5.21}
\end{equation}
Typical shapes of the solitary waves are presented in Fig. \ref{sol_wave_fig}. Figure \ref{sol_wave_fig} shows that
the only type of possible solitary waves are depression waves. It is known that the depression capillary-gravity solitary waves
are possible \citep[see][]{Korteweg de Vries,Benjamin1982,Vanden-Broeck1983} and have been observed experimentally in fluids
with sufficiently strong surface tension \citep[][]{Falcon}. This
gives us one more argument in favour of our earlier conclusion that the effect of vibrations is similar to that of surface tension.

\begin{figure}[h]
\begin{center}
\includegraphics*[height=10cm]{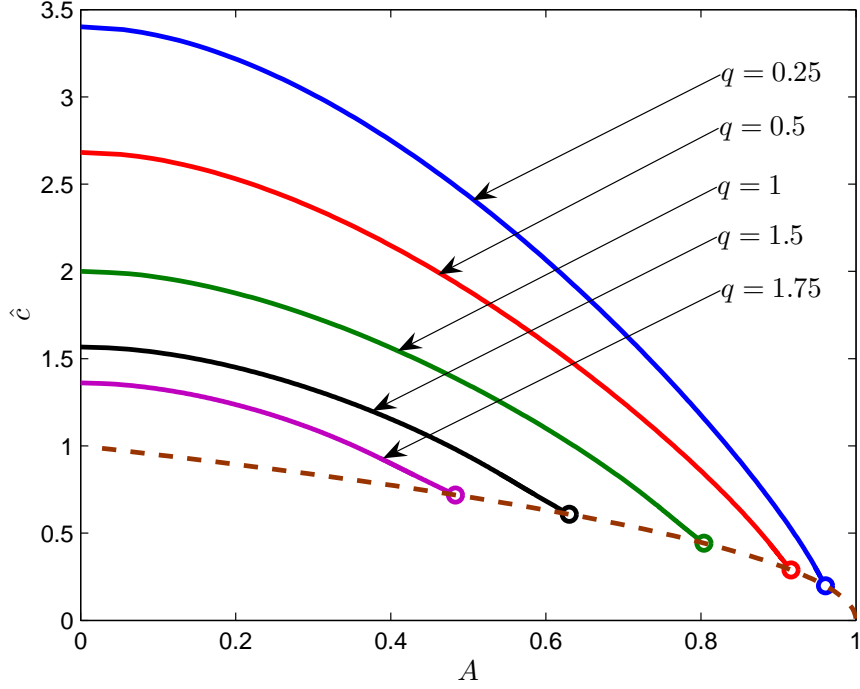}
\end{center}
\caption{Wave speed $\hat{c}=c/\sqrt{\gamma_0}$ versus amplitude $A$. Solid curves show the wave speed of periodic waves for several values of parameter $q$. The dashed curve corresponds to solitary waves. Circles represent the limit points of the wave speed of periodic waves when the wavelength goes to infinity.}
\label{wave_speed_fig}
\end{figure}

The wave speed, $\hat{c}=c/\sqrt{\gamma_0}$, versus the amplitude, $A$ in shown in Fig. \ref{wave_speed_fig} as a dashed curve. It approaches
$1$ when the amplitude decreases to zero and $0$ when the amplitude approaches $1$. The solid curves in Fig. \ref{wave_speed_fig} correspond to periodic
travelling waves discussed below.

\subsection{Periodic waves}

It follows from Fig. \ref{potential} that the motion of the particle is periodic if
the particle's energy, $\hat{E} = R'^{2}/2 + \hat{V}(R)$, satisfies the inequality
\begin{equation}
\hat{V}(R_1)< \hat{E} < \hat{V}(R_2). \label{5.22}
\end{equation}
Periodic motions of the particle correspond to periodic travelling waves solutions of the original equations (\ref{5.1}) and (\ref{5.2}).
For each $q\in(0,2)$, there is a family of periodic solutions corresponding to values of the particle's energy, $\hat{E}$, sattisfyilng
inequality (\ref{5.22}). In what follows, we restrict our attention to waves for which the mass flux (equivalently, the momentum density),
$M=HU$, averaged over
the period of the wave, is zero, i.e.
\begin{equation}
\langle M \rangle \equiv \frac{1}{S} \, \int\limits_{0}^{S} H(s)U(s) \, ds = 0. \label{5.23}
\end{equation}
Here $S$ is the period (wavelength) of the wave. Also, to be consistent with our non-dimentionalisation, we require that
the averaged (over the period) fluid depth is equal to $1$, i.e.
\begin{equation}
\langle H \rangle \equiv \frac{1}{S} \, \int\limits_{0}^{S} H(s) \, ds = 1. \label{5.24}
\end{equation}
On averaging Eq. (\ref{5.8}) over the period and taking into account Eqs. (\ref{5.23}) and (\ref{5.24}), we  find the relationship between
$B$ and $c$:
\begin{equation}
B=-c \quad \hbox{or} \quad \hat{B}=-\hat{c} . \label{5.25}
\end{equation}
Since the particle's energy is a constant of motion ($d\hat{E}/ds=0$), we have
\begin{equation}
R'(s)=\pm \, \sqrt{2\left(\hat{E}-\hat{V}(R)\right)}. \label{5.26}
\end{equation}
Hence, the period of motion of the particle is given by
\begin{equation}
\Sigma=2 \, \int\limits_{R_{min}}^{R_{\max}} \frac{dR}{\sqrt{2(\hat{E}-\hat{V}(R))}} \label{5.27}
\end{equation}
where $R_{max}$ and $R_{\min}$ are the maximum and minimum values of $R$ corresponding to periodic motion of the particle with energy
$\hat{E}$ (in other words, $R_{max}$ and $R_{\min}$ are solutions of the equation $\hat{V}(R)=\hat{E}$). In view of (\ref{5.12}),
the period of function $H(s)$
(the wavelength) is then given by
\begin{equation}
S=m^{1/2}\left(\frac{\hat{C}}{3}\right)^{1/2} \, \Sigma. \label{5.28}
\end{equation}
Further, on rewriting the integral in Eq. (\ref{5.24}) in terms of $R$ and $\sigma$, we obtain
\begin{equation}
\left(\frac{\hat{C}}{3}\right)^{-1} = \frac{1}{\Sigma} \, \int\limits_{0}^{\Sigma}R^{2/3}(\sigma) \, d\sigma
= \frac{2}{\Sigma} \, \int\limits_{R_{min}}^{R_{\max}} \frac{R^{2/3} \, dR}{\sqrt{2(\hat{E}-\hat{V}(R))}} . \label{5.29}
\end{equation}
Now all properties of the periodic solutions can be found as follows.
First, we fix $q\in(0,2)$. Then for each $\hat{E}$ satisfying (\ref{5.22}),
we compute $\hat{C}$, $S$ (using (\ref{5.29}) and (\ref{5.28}), respectively) and $\hat{c}$ (using Eqs. (\ref{5.15}) and (\ref{5.25})), as well as
the wave amplitude, $A$, defined by
\begin{equation}
A=\max_{s\in[0,S]}\vert H(s)-1\vert = \frac{\hat{C}}{3} \, \max \left\{1-R_{min}^{2/3}, R_{max}^{2/3}-1\right\}.  \label{5.30}
\end{equation}
As a result, we get the period, $S$, and the wave speed, $\hat{c}$, as functions of the amplitude, $A$.
\begin{figure}
\begin{center}
\includegraphics*[height=10cm]{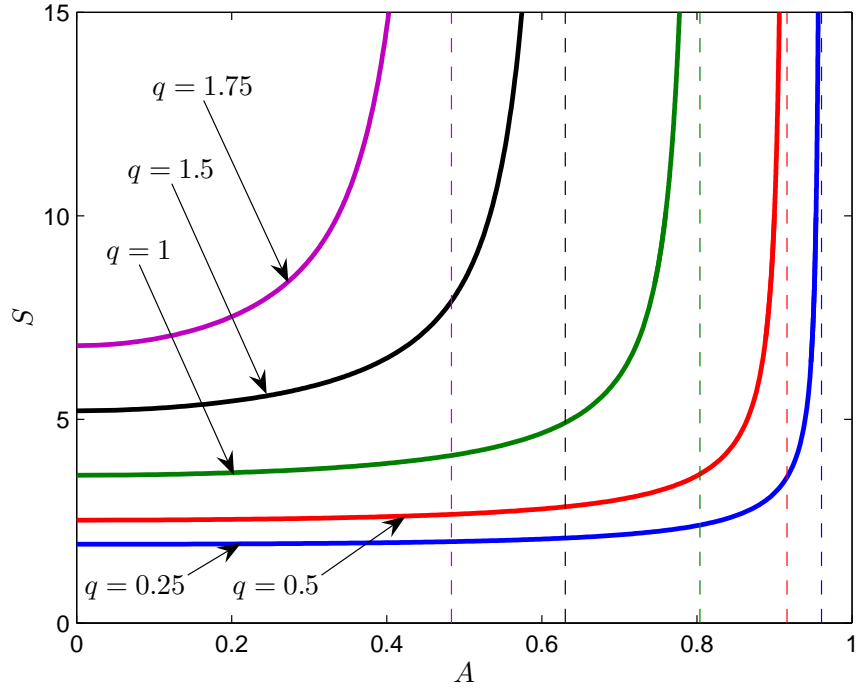}
\end{center}
\caption{Period $S$ versus amplitude $A$ for several values of parameter $q$. Dashed vertical lines represent the asymptotes of the curves $S(A)$.
These asymptotes correspond to solitary waves.}
\label{period_all_q_fig_NEW}
\end{figure}

\begin{figure}
\begin{center}
\includegraphics*[height=10cm]{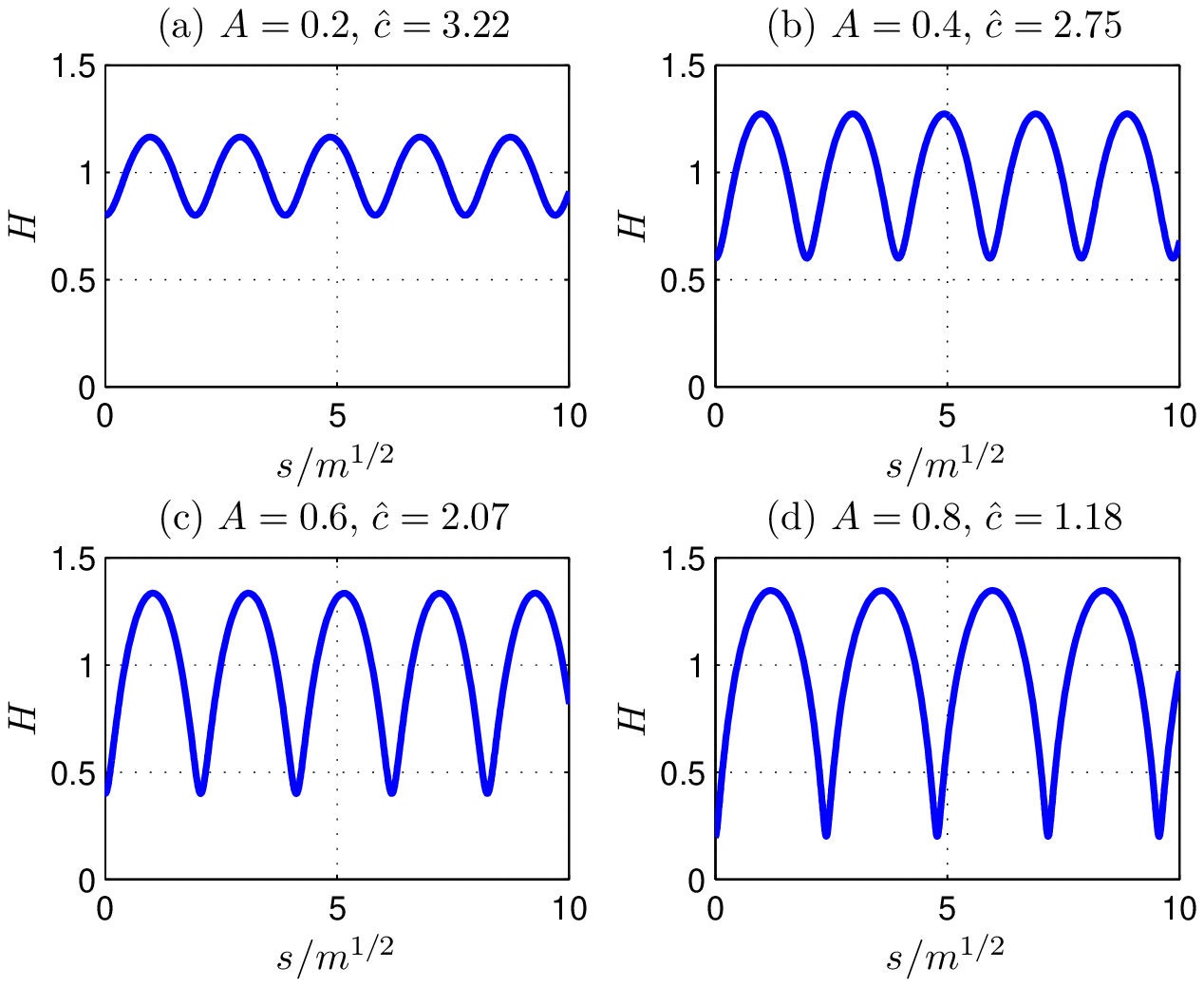}
\end{center}
\caption{Periodic waves for $q=0.25$ and several values of the amplitude $A=\max \vert H(s) - 1\vert$; $\hat{c}=c/\sqrt{\gamma_0}$ where
$c$ is the wave speed.}
\label{periodic_waves_fig1}
\end{figure}

\begin{figure}
\begin{center}
\includegraphics*[height=10cm]{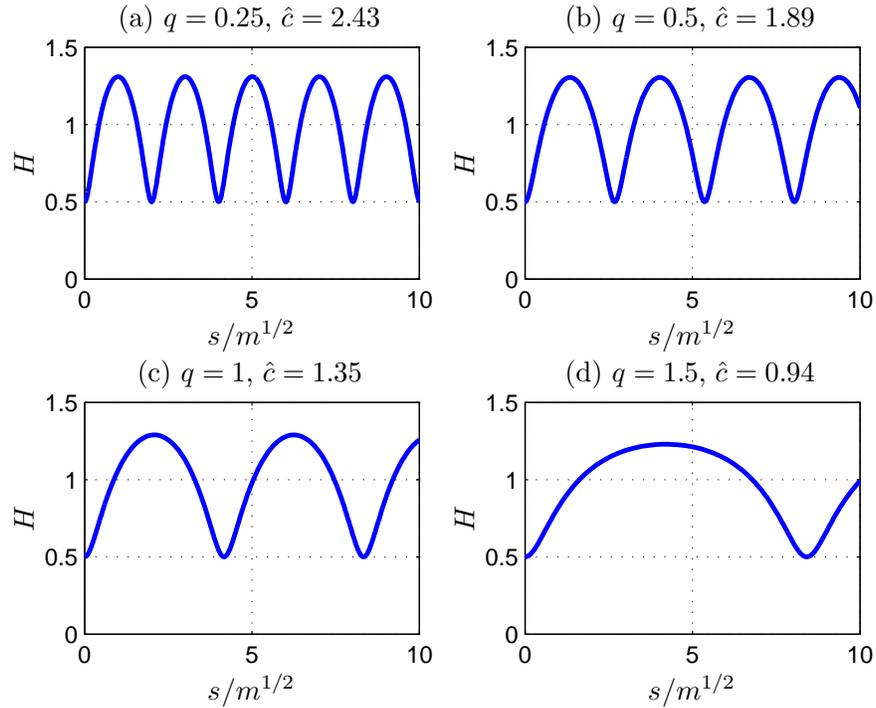}
\end{center}
\caption{Periodic waves of the same amplitude $A=0.5$ for several values of parameter $q$.}
\label{periodic_waves_fig2}
\end{figure}

The period versus the amplitude for several values of parameter $q$ are shown in Fig. \ref{period_all_q_fig_NEW}.
The figure shows that
for each value of $q$, $S(A)$ increases with $A$ and goes to infinity as $A$ approaches a certain critical value, which corresponds
to the amplitude of the solitary wave for the same value of $q$. The dashed vertical asymptotes of the curves $S(A)$, shown in
Fig. \ref{period_all_q_fig_NEW}, represent the amplitudes of the corresponding solitary waves. The wave speed as a function of the amplitude
for several values of $q$ is shown in Fig. \ref{wave_speed_fig}: solid curves. For each $q$, $\hat{c}$ decreases as $A$ varies from $0$
up to the critical amplitude (corresponding to a solitary wave). The end points of the curves, $\hat{c}(A)$ are are indicated by circles
in Fig. \ref{wave_speed_fig}. Examples of the wave shapes for various values of $q$ and $A$ are shown in Figs. \ref{periodic_waves_fig1}
and \ref{periodic_waves_fig2}.

\subsection{Very strong vibration ($\gamma_0=0$)}

\begin{figure}
\begin{center}
\includegraphics*[height=10cm]{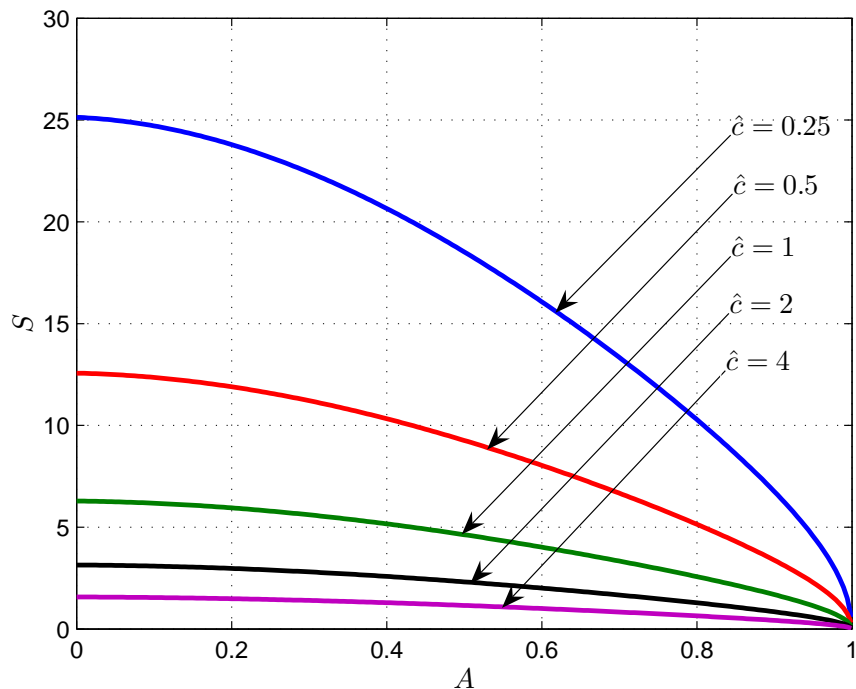}
\end{center}
\caption{Periodic travelling waves for $\gamma_0=0$: the wavelength, $S$, versus the amplitude,
$A$, for several values of the wave speed.}
\label{limit_S_of_Amp_fig}
\end{figure}

In the case of $\gamma_0=0$, Eq. (\ref{5.10}) can be rewritten as
\begin{equation}
\left(H \, H''+\frac{H'^2}{2}\right)= \frac{B^{*2}}{2H^2}+H - C^*  \label{5.31}
\end{equation}
where (cf. (\ref{5.11}))
\begin{equation}
B^*=\frac{B}{\sqrt{\varkappa}}, \quad C^*=\frac{C}{\varkappa}+\frac{c^{*2}}{2},
 \quad c^*=\frac{c}{\sqrt{\varkappa}}. \label{5.32}
\end{equation}
In terms of new variables $R$ and $\sigma$, defined as (cf. (\ref{5.12}))
\begin{equation}
H(s)=\sqrt{\frac{B^{*2}}{2C^*}} \, R^{2/3}(\sigma), \quad \sigma = \sqrt{\frac{2C^{*2}}{B^{*2}}} \, s, \label{5.33}
\end{equation}
Eq. (\ref{5.31}) reduces to the equation of motion of a particle in a potential:
\begin{equation}
R''(\sigma) = - \hat{V}'(R), \quad \hat{V}(R) = \frac{9}{4} \, \left( R^{-2/3} +  R^{2/3}\right).  \label{5.34}
\end{equation}
Note that the potential $\hat{V}(R)$ does not contain free parameters. It has a global minimum at $R=1$, so that
only periodic motion of the particle is possible.
As before, we restrict our attention to waves for which the mass flux, $M=HU$, averaged over
the period of the wave, is zero ($\langle M\rangle=0$) and require that the mean depth is equal to $1$ ($\langle H\rangle=1$).
Let
\begin{equation}
\lambda=\sqrt{\left(2\hat{E}/9\right)^2-1} \label{5.35}
\end{equation}
where $\hat{E}=R'^2+\hat{V}(R)$ is the particle's energy. Periodic solutions are possible for $9/2=\hat{V}(1) < \hat{E} < \infty$. This implies that
parameter $\lambda$, defined by (\ref{5.35}), can vary from $0$ to $\infty$.
The equation of motion (\ref{5.34}) can be integrated analytically (albeit in implicit form). We will present only final results.
It can be shown that the period, $S$, and the amplitude, $A$, can be expressed in terms of $\lambda$ as
\begin{equation}
S= \frac{2\pi}{c^*} \, \frac{\left(1+\lambda^2\right)^{3/2}}{\left(1+\frac{3}{2} \, \lambda^2\right)^2}, \quad
A= \frac{\lambda\sqrt{1+\lambda^2}+\frac{1}{2} \, \lambda^2}{1+\frac{3}{2} \, \lambda^2}. \label{5.36}
\end{equation}
The wave amplitude does not depend on the wave speed. It goes to $0$ as $\lambda\to 0$ and to $1$ when $\lambda\to\infty$.
The period (wavelength) is inversely proportional to the wave speed and tends to a nonzero limit, depending on $c^*$,
as $\lambda\to 0$ and goes to $0$ when $\lambda\to\infty$. For each $c^*$, Eq. (\ref{5.36}) defines a curve on the $(A,S)$-plane.
These curves for several values of the wave speed are shown in Fig. (\ref{limit_S_of_Amp_fig}).
%%%%%%%%%%%%%%%%%%%%%%%%%%%%%%%%%%%%%%%%%%%%%%%%%%%%%%%%%%%%%%%%%%%%%%%%%%%%%%%%%%%%%%%%%%%%%%%%%%%%%%%%%%%%%%%%%%%%%%%%%%%%%%%%%%%%

\section{Discussion}

We have considered  free-surface waves in a layer of an inviscid fluid which is subject to vertical vibrations
and derived three asymptotic models for slowly evolving nonlinear waves in the limit of high-frequency
vibrations. This has been done by constructing asymptotic expansions of the exact governing equations for water waves
using the method of multiple scales.
All three expansion resulted in asymptotic equations that are nonlinear and valid for waves whose amplitude is
of the same order of magnitude as  the fluid depth and for vibrations of sufficiently high frequency such that the acceleration
due to vibrations is much greater than the gravitational acceleration. In the first expansion, it was also assumed that
the amplitude of the vibration is small (relative to the fluid depth). The asymptotic equations that emerge in this case
are Hamiltonian and coincide with the standard equations for water waves in a non-vibrating
fluid layer with an additional non-local nonlinear term representing the effect of the vibration. This leads to an extra term
in the dispersion relation for linear waves, which, at least in the long-wave limit, is similar to what would happen is
a surface tension was taken into account. The second and the third expansions employed the long-wave approximation. In the second
one, it was assumed that the vibration amplitude was small, while in the third
it was of the same order of magnitude as the fluid depth. Remarkably, these two quite different physical assumptions lead to
the nonlinear equations that have exactly the same form. Again, the equations are Hamiltonian.
It the vibration is absent, they reduce to the standard dispersionless
shallow water equations. The effect of the vibration appears as an extra nonlinear term which makes the equations
dispersive and, for linear waves, is equivalent
to surface tension.

The analysis of one-dimensional waves has shown that the asymptotic equations have travelling solitary and periodic wave solutions.
The solitary waves are depression waves and have wave speed smaller than that of gravity waves (subcritical waves).
This resembles the capillary-gravity depression waves that were predicted first by \citet{{Korteweg de Vries}}, later
considered again by \citet{Benjamin1982} and \citet{Vanden-Broeck1983} and recently observed in experiments by \citet{{Falcon}}.
Our equations have neither solution in the form of elevation solitary waves, nor solitary wave solutions propagating with the speed
higher than that of gravity waves
(which is different from the capillary-gravity waves in a non-vibration fluid where such waves exist). At the moment,
the reason for the non-existence of elevation waves it is not quite clear: it may be related to the approximate
nature of the model equations or it may be a generic property of the waves in vibrating fluid. This question requires a further
investigation.

We have also shown that the asymptotic equations have many periodic travelling wave solution and that
the solitary waves can be viewed as the limit of periodic waves as the wavelength of the latter
is continuously increased. It has  also been shown that if the vibration completely dominates over the gravity,
the travelling wave solutions of the asymptotic equations can be found analytically and that only
periodic waves are possible in this case.

Here we did not consider weakly nonlinear waves in the vibrating fluid layer. An interesting open question that arises in this context is
whether there are slowly varying waves that can be described by the standard KdV equation or some other equation will emerge.
This is a subject of a continuing investigation.

%%%%%%%%%%%%%%%%%%%%%%%%%%%%%%%%%%%%%%%%%%%%%%%%%%%%%%%%%%%%%%%%%%%%%%%%%%%%%%%%%%%%%%%%%%%%%%%%%%%%%%%%%%%%%%%%%%%%%%%%%%%%%%%%%%%%
\vskip 5mm
\textbf{Acknowledgement.} The author is grateful to Profs. S. L. Gavrilyuk, A. B. Morgulis and M. Yu. Zhukov and  for interesting
and useful discussions.

%%%%%%%%%%%%%%%%%%%%%%%%%%%%%%%%%%%%%%%%%%%%%%%%%%%%%%%%%%%%%%%%%%%%%%%%%%%%%%%%%%%%%%%%%%%%%%%%%%%%%%%%%%%%%%%%%%%%%%%%%%%%%%%%%%%%

\bibliographystyle{jfm}

\end{document}